\documentclass[preprint,showpacs,preprintnumbers,amsmath,amssymb]{revtex4}
\usepackage{graphicx}
\usepackage{epsfig}
\usepackage{bm}
\usepackage{amsfonts}
\usepackage{subfigure}
\usepackage{multirow}
\usepackage{float}
\usepackage{color}
\usepackage{ulem}
\usepackage{todonotes}

\usepackage{makecell}
\usepackage{enumitem}
\usepackage{float}
\usepackage{amsmath}
\usepackage{hyperref}
\hypersetup{colorlinks=True,citecolor=blue}
\usepackage{amsfonts}
\usepackage{amssymb}
\usepackage{caption}
\usepackage{float}
\usepackage{bm}
\usepackage{graphicx}
\usepackage{color}
\usepackage{verbatim}
\usepackage{subfigure}
\usepackage{multirow}

\def\thickhline{\noalign{\hrule height1.5pt}}

\begin{document}

\title{Mutated hilltop inflation in light of Planck/ACT observations}
\author{Iraj Safaei$^{1}$\footnote{isafaei@kashanu.ac.ir}, Soma Heydari$^{1}$\footnote{s.heydari@uok.ac.ir}, Milad Solbi$^{2}$\footnote{miladsolbi@gmail.com} and Kayoomars Karami$^{1}$\footnote{kkarami@uok.ac.ir}}

\affiliation{\small{
$^{1}$Department of Physics, University of Kurdistan, Pasdaran Street, P.O. Box 66177-15175, Sanandaj, Iran\\
$^{2}$Khon Kaen Particle Physics and Cosmology Theory Group (KKPaCT), Department of Physics, Faculty of Science, Khon Kaen University, 123 Mitraphap Rd., Khon Kaen 40002, Thailand
}}


\begin{abstract}
Here, a single field inflationary model driven by a mutated hilltop potential, a subclass of the hilltop models of inflation, is investigated. To constrain the parameter space, we employ the latest $r-n_{\rm s}$ constraints from Planck 2018, BICEP/Keck 2018, and the Atacama Cosmology Telescope (ACT) data, alongside reheating parameters  $N_{\rm{re}}$, $T_{\rm{re}}$, and  $\omega_{\rm{re}}$, and the model independent bound on the radiation dominated (RD) era $N_{\rm{rd}}$. Furthermore,  the relic gravitational wave (GW) spectrum within the sensitivity domains of future GW detectors are analyzed. By combining CMB, reheating, RD era, and  GW constraints, we find for the Planck+BK18 data that the inflationary duration is confined to $46 \leq N \leq 56$ (95\% CL) and $48.1 \leq N \leq 56$ (68\% CL). Moreover, the model parameter $\alpha$ is confined to $0.161 \leq \alpha \leq 0.890$ (95\% CL) and $0.217 \leq \alpha \leq 0.815$ (68\% CL). Inclusion of the ACT data further tighten the constraints to $54 \leq N \leq 56$ (95\% CL) and $0.29 \leq \alpha \leq 0.62$ (95\% CL), thereby enhancing the precision and robustness of the model predictions.
\end{abstract}
\maketitle
%
%
\section{Introduction}

Inflation provides a compelling resolution to the horizon, flatness, and monopole problems of the standard Big Bang cosmology \cite{Starobinsky:1980,Guth:1981, Linde:1982}. During the inflation, quantum fluctuations of the inflaton field generate the scalar and tensor perturbations, which imprint the anisotropies observed in the cosmic microwave background (CMB). The first step in assessing any inflationary model is to compare its predictions for the scalar spectral index $n_s$ and the  tensor-to-scalar ratio $r$ with CMB observation, thereby constraining the model parameters \cite{Eshaghi:2016,Mishra:2021,safaei:2024,Karčiauskas:2022}. Recent measurements from the Atacama Cosmology Telescope (ACT) \cite{ACT2025} have provided improved constraints on the scalar spectral index $n_{\rm s}$ that offer complementary precision to Planck and BICEP/Keck data \cite{akrami:2020,bk18}. The inclusion of ACT measurements, particularly in combination with Planck 2018 and BK18, enables a more stringent test of inflationary models that reduce the viable parameter space and enhance the robustness of our analysis. In the next step, one can apply the constraints on post inflationary parameters, including those associated with the reheating phase, the radiation dominated (RD) era, and relic gravitational waves (GWs) to further refine the parameter space deduced from the CMB data.

Most inflationary models feature a potential with a flat region to sustain a slow-roll  evolution of the inflaton field.
After the end of inflation, the inflaton rolls down the potential and enters an
oscillatory phase around the potential minimum. This oscillatory phase,   known as the reheating epoch, connects the inflationary era to the RD era \cite{Kofman:1994,Martin:2010,Unnikrishnan:2012,LiangDai:2014,Cook:2015,Gialamas:2020,sahni:1990,Wolf:2024}. During reheating, the inflaton decays into standard model particles that results in thermalizing the Universe. The reheating phase is characterized by three parameters such as the duration $N_{\rm reh}$, the temperature $T_{\rm re}$, and the equation of state parameter $\omega_{\rm re}$.
 These parameters are often used to further constrain  inflationary model. Following the reheating phase, the RD era begins in the evolution of the Universe. The reheating parameters influence the duration of the RD era $N_{\rm rd}$. Hence, model independent bounds on the duration of the RD era provide additional constraints on the parameter space of the inflationary models \cite{German:2023}.

A further method to constraining the inflationary models is to employ relic GWs \cite{Mishra:2021,safaei:2024}. The relic GWs are generated from quantum fluctuations of the inflaton field during the inflationary era. They  propagate through space-time and carry valuable information about  the physics of the early Universe, owing to their non-interacting essence. The density parameter spectrum of these relic GWs is affected by the post inflationary equation of state parameter $\omega_{\rm{re}}$ \cite{sahni:1990}, which, consequently, depends on the model parameters. Therefore, the relic GWs serve as a powerful tool for further refining the parameter space obtained from the implications of the reheating and RD eras.

Among the multifarious inflationary models, hilltop inflation has drawn significant interest from the scientific community, due to its compatibility with small field inflation scenarios, that are in consistency with observations \cite{Cook:2015,Wolf:2024}. In recent years, variants of hilltop inflation with non-polynomial modifications, such as the mutated hilltop model, have been introduced to enhance the pliability and applicability of this framework. The mutated hilltop model, rooted in supergravity \cite{Pinhero:2019},  incorporates a scalar potential with a hyperbolic secant term. In this model, transitions between small field and large field inflationary regimes are modulated by the parameter $\alpha$ \cite{Pal:2010,Kumar:2018, Yadav:2024}. This tunable characteristic makes the mutated hilltop model  suitable for exploring observational constraints on the inflationary parameter space. In this model, the inflaton evolves from the hilltop of the potential toward its minimum around $\phi=0$, driving an observable inflationary era. Notably, the scalar spectral index $n_s$ predicted by the mutated hilltop potential remains largely insensitive to the model parameter, while the tensor-to-scalar ratio $r$ varies over a range of $10^{-4}\leq r\leq 10^{-1}$, depending on the choice of parameters \cite{Kumar:2018}. This range confirms the model versatility in addressing a wide spectrum of inflationary predictions.

The aim of this study is to examine the observational constraints on the mutated hilltop inflationary potential with a special focus on the role of the model parameter $\alpha$ in shaping inflationary observables. Additionally, we study the implications of the reheating and RD phases for the model predictions in order to analyzing how  their key  parameters $N_{\rm re}$, $T_{\rm re}$, $\omega_{\rm re}$ and $N_{\rm rd}$ vary as functions of $\alpha$. Furthermore, we assess the energy density spectrum of  relic GWs and their detectability by upcoming GW observatories like BBO \cite{Yagi:2011BBODECIGO,Yagi:2017BBODECIGO,Crowder:2005BBO,Corbin:2006BBO,Harry:2006BBO},
DECIGO \cite{Yagi:2011BBODECIGO,Yagi:2017BBODECIGO,Kawamura:2006DECIGO,Kawamura:2011DECIGO,Seto:2001DECIGO},
LISA \cite{lisa,lisa-a},
SKA \cite{skaCarilli:2004,ska,skaWeltman:2020}, and
PTA \cite{epta1:add,epta2:add,epta3:add,epta4:add,epta5:add}, providing insights into the observational prospects for testing the predictions of the mutated hilltop model.

This paper is classified as follows: In Section \ref{sec2},  the basic outline of the mutated hilltop inflation is introduced. Section \ref{sec3} is devoted to applying  the reheating and RD constraints to the model. In Section \ref{sec4}, the prediction of the model for  relic GWs is discussed. Finally, the main conclusions of the paper are summarized in Section \ref{sec5}.


\section{Mutated hilltop inflation}\label{sec2}
The action for the present model is given by \cite{Guth:1981,Linde:1982}
\begin{equation}\label{eq:action}
S=\int{\rm d}^{4}x \sqrt{-g} \left[\frac{R}{16\pi G}+{\cal L}(X,\phi)\right] ,
\end{equation}
where $g$ and $R$ denote the determinant of the metric tensor $g_{\mu\nu}$ and the Ricci scalar, respectively.
The term $\mathcal{L}(X,\phi)$ represents the Lagrangian density, where $\phi$ is the scalar field, and  $X\equiv\frac{1}{2}g_{\mu\nu}~\partial^\mu \phi \partial^\nu \phi$ is the kinetic energy term.
Here, we consider a canonical Lagrangian given by
\begin{equation}\label{Lagrangian}
{\cal L}(X,\phi) = X - V(\phi) ,
\end{equation}
where $V(\phi)$ is the scalar field potential.
We assume the spatially flat Friedmann-Robertson-Walker (FRW) metric for $g_{\mu\nu}$ as
\begin{equation}
\label{eq:FRW}
{\rm d} {s^2} = {\rm d} {t^2} - {a^2}(t)\left( {{\rm d} {x^2} +
{\rm d} {y^2} + {\rm d} {z^2}} \right) ,
\end{equation}
where $a(t)$ is the scale factor and  $t$ is the  cosmic time. Thus,  the kinetic term  simplifies to $X=\dot{\phi}/2$. Additionally,
the energy density $\rho_{\phi}$ and pressure $p_{\phi}$ of the scalar field corresponding to the Lagrangian (\ref{Lagrangian}) are given by
\begin{equation}
\rho _\phi =\frac{1}{2}\dot{\phi }^{2} + V(\phi) \label{eq:rho},
\end{equation}
\begin{equation}
{p_\phi } = \frac{1}{2}\dot{\phi }^{2}- V(\phi) \label{eq:Lp}.
\end{equation}
Using Eqs. (\ref{eq:rho}) and (\ref{eq:Lp}), the equation of state parameter is defined as
$\omega _{\phi }\equiv {p_{\phi }}/{\rho _{\phi }}$.
Furthermore,  taking the variation of the action  (\ref{eq:action}) with respect to the metric (\ref{eq:FRW})  yields the first  and second Friedmann equations as follows
\begin{eqnarray}
&H^{2} =\dfrac{1}{3M_{\rm p}^2} \rho_{\phi} , \label{eq: FR-eqn1} \\
&\dot{H} = -\dfrac{1}{2M_{\rm p}^2} (\rho _{\phi} +p_{\phi} )  ,
\label{eq:FR-eqn2}
\end{eqnarray}
where $H \equiv \dot{a}/a$ is the Hubble parameter, $M_{\rm p} \equiv 1 / {\sqrt{8\pi G}}$ is the reduced Planck mass, and the dot denotes the derivative with respect to $t$.
Moreover, the equation of motion for the scalar field, derived by varying the action (\ref{eq:action}) with respect to $\phi$, is given by
\begin{equation}
\label{eq:KG}
\ddot{\phi }+3H\dot{\phi }+V'(\phi )=0,
\end{equation}
where $({}')$ indicates the derivative with respect to $\phi$. In what follows,
the first and second slow-roll parameters are defined as
\begin{equation}\label{eq:epsilon 1}
\epsilon_{\rm H} \equiv - \frac{\dot{H}}{H^2},
\end{equation}
\begin{equation}\label{eq:eta 1}
\eta_{\rm H} \equiv - \frac{\ddot{\phi}}{H \dot{\phi}}.
\end{equation}
During the slow-roll phase,  the conditions  $(\epsilon_H , \eta_H) \ll 1$ are satisfied.
Under the slow-roll conditions, the background Eqs. (\ref{eq: FR-eqn1}) and (\ref{eq:KG}) can be simplified to
\begin{eqnarray}
& H^2 \simeq \dfrac{1}{3M_p^2}~V(\phi ), \label{eq:H2}\\
& 3H\dot{\phi }+V'(\phi ) \simeq 0. \label{eq:KG SR}
\end{eqnarray}
Consequently, the slow-roll parameters can be rewritten as
\begin{equation}\label{eq:epsilon H}
\epsilon _{\rm H}(\phi) \simeq 2M_{\rm p}^2\left ( \frac{H'\left ( \phi  \right )}{H\left ( \phi  \right )} \right )^2,
\end{equation}
\begin{equation}\label{eq:eta H}
\eta _{\rm H}(\phi) \simeq 2M_{\rm p}^2 \left ( \frac{H'' \left ( \phi  \right )}{H\left ( \phi  \right )} \right ).
\end{equation}

It is well recognized that the evolution of scalar and tensor perturbations can be characterized by their respective power spectra denoted by ${\cal P}_{\rm s}$ and ${\cal P}_{\rm t}$. Under the slow-roll conditions, they can be approximated at the time of horizon exit $k=aH$ as follows
\begin{eqnarray}
&{\cal P}_{\rm s}\simeq\dfrac{H^2}{8 \pi ^{2}M_{\rm p}^{2} \epsilon _{\rm H}}\Big|_{k=aH},\label{eq:Ps-SR} \\
&{\cal P}_{\rm t}\simeq\dfrac{2H^2}{\pi ^{2}M_{\rm p}^2}\Big|_{k=aH} \label{eq:Pt-SR}.
\end{eqnarray}
The value of the scalar power spectrum at the pivot scale $k_\ast=0.05~{\rm Mpc^{-1}}$ has been estimated by Planck measurements as  ${\cal P}_{\rm s}(k_*)=2.1\times10^{-9}$ \cite{akrami:2020}.
Utilizing Eqs. (\ref{eq:epsilon H})-(\ref{eq:Pt-SR}), the scalar and tensor spectral indices in terms of the slow-roll parameters are calculated as
\begin{eqnarray}
& n_{\rm s} - 1\equiv \dfrac{\mathrm{d} \ln {\cal P}_ {\rm s}}{\mathrm{d} \ln k}=-4\epsilon _{\rm H}+2\eta _{\rm H}, \label{eq:inf_ns_SR} \\
& n_{\rm t}\equiv \dfrac{\mathrm{d} \ln {\cal P}_{\rm t}}{\mathrm{d} \ln k}= -2 \epsilon _{\rm H}. \label{eq:ntt}
\end{eqnarray}
In addition, the tensor-to-scalar ratio is given by
\begin{equation}
\label{eq:r}
r\equiv\frac{{\cal P}_{\rm t}}{{\cal P}_{\rm s}}= 16  \epsilon _{\rm H}.
\end{equation}
Note that from Eqs. (\ref{eq:ntt}) and (\ref{eq:r}), the consistency relation is obtained as
\begin{equation}
r=-8n_{\rm t}.
\end{equation}
The Planck 2018 TT, TE, EE + LowE + Lensing + BK18 + BAO measurements impose the following constraints on $n_{\rm s}$ and $r$ \cite{akrami:2020,bk18}
\begin{eqnarray}
&n_{\rm s} = 0.9653_{-\,0.0041\,-\,0.0083}^{+\,0.0041\,+\,0.0107}, \nonumber \\
&r < 0.036. \label{eq:nsr95}
\end{eqnarray}
The inclusion of the ACT measurements further refines the constraints on $n_{\rm s}$ as follows  \cite{ACT2025}
\begin{equation} \label{eq:nsr95ACT}
n_{\rm s} = 0.9743_{-\,0.0034\,-\,0.0078}^{+\,0.0034\,+\,0.0077}.
\end{equation}
Here, we focus on the mutated hilltop potential for our model in light of the latest observational constraints. This potential can be considered as a variant of the broader, well-known hilltop inflationary potentials. The term mutated indicates that the potential deviates from the standard hilltop form and incorporates features from hybrid inflation scenarios. In this model, inflation occurs as the inflaton rolls down the hilltop of the potential, located at small values of $\phi$. The mutated hilltop potential is known to drive a more consistent inflationary era with observational data compared to the standard hilltop potentials.
The mutated hilltop potential is given by \cite{Pal:2010,Kumar:2018, Yadav:2024}
\begin{equation}\label{eq:MHI potential}
V\left ( \phi  \right )=V_0\left [ 1-{\rm sech} \left ( \alpha \phi   \right ) \right ] ,
\end{equation}
where $V_0$ and  $\alpha$ are constant parameters with dimensions of $M^4$ and $M^{-1}$, respectively.
Using Eqs. (\ref{eq:H2}) and (\ref{eq:MHI potential}), the slow-roll parameters (\ref{eq:epsilon H}) and (\ref{eq:eta H}) take the following forms
%
\begin{eqnarray}
&\epsilon _{\rm H}(\phi)\simeq \dfrac{{M_{\rm p}}^2 \alpha^2~ {\rm sech}^2( \alpha  \phi)\tanh ^2( \alpha \phi )}{2\big[ 1-{\rm sech}( \alpha \phi) \big]^2}, \label{eq:epsilon} \\
&\eta_{\rm H}(\phi)\simeq \dfrac{-{M_{\rm p}}^2 \alpha^2~{\rm sech}( \alpha \phi)}{2}~\big[ 2+3~{\rm sech}( \alpha \phi)\big]. \label{eq:eta}
\end{eqnarray}
Within the slow-roll approximation, inflation ends when one of the slow-roll conditions is violated.
In this model, it has been demonstrated that for $\alpha > \alpha_{\rm eq}$, $|\eta_{\rm H}|$ exceeds one before $\epsilon_{\rm H}$.
Consequently, the end of inflation is determined by the condition where either $\epsilon_{\rm H} =1$ or $|\eta_{\rm H}|=1$ is first satisfied \cite{Kumar:2018}.
Here,  $\alpha_{\rm eq}$ indicates the values of $\alpha$ at which $\epsilon_{\rm H} =1$ and $|\eta_{\rm H}|=1$ are simultaneously satisfied.
Note that the scalar field value at the end of inflation serves as a boundary condition for solving the background equation (\ref{eq:KG SR}).

\begin{figure}[H]
\begin{minipage}[b]{1\textwidth}
\vspace{-.1cm}
\centering
\subfigure[\label{phiN}]{\includegraphics[width=.48\textwidth]%
{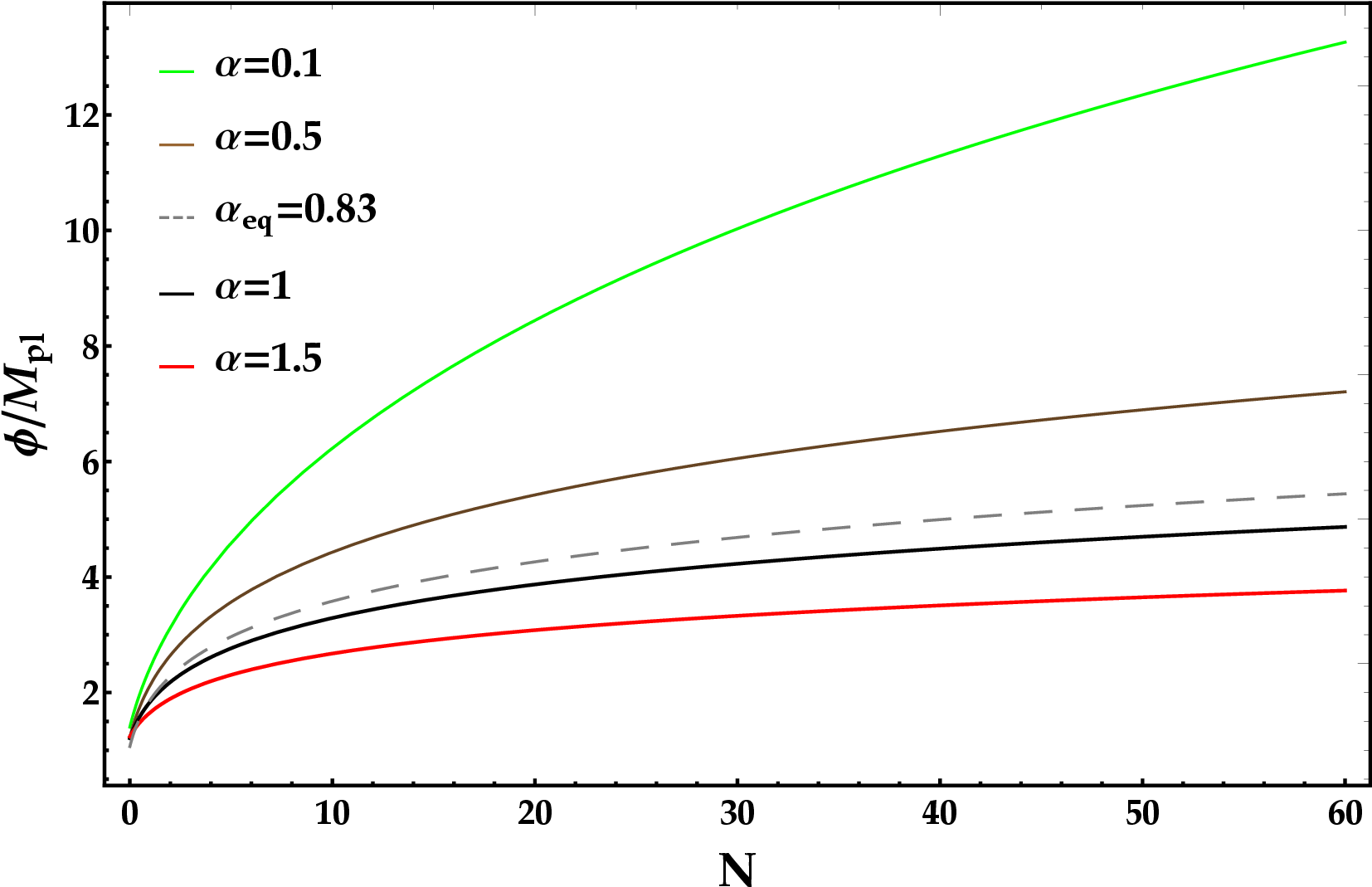}} \hspace{.1cm}
\centering
\subfigure[\label{HN}]{\includegraphics[width=.48\textwidth]%
{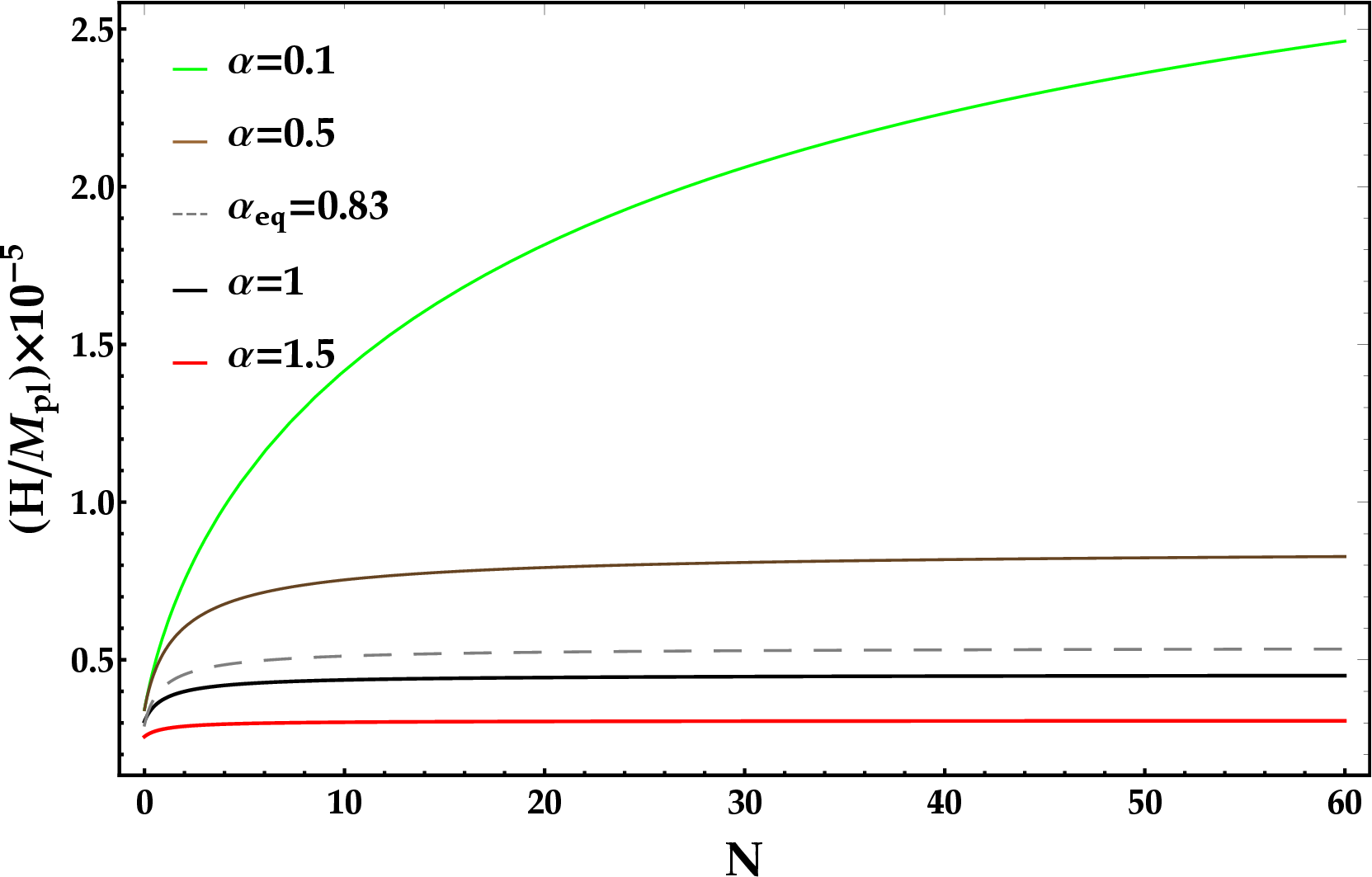}} \hspace{.1cm}
\centering
\subfigure[\label{eN}]{\includegraphics[width=.48\textwidth]%
{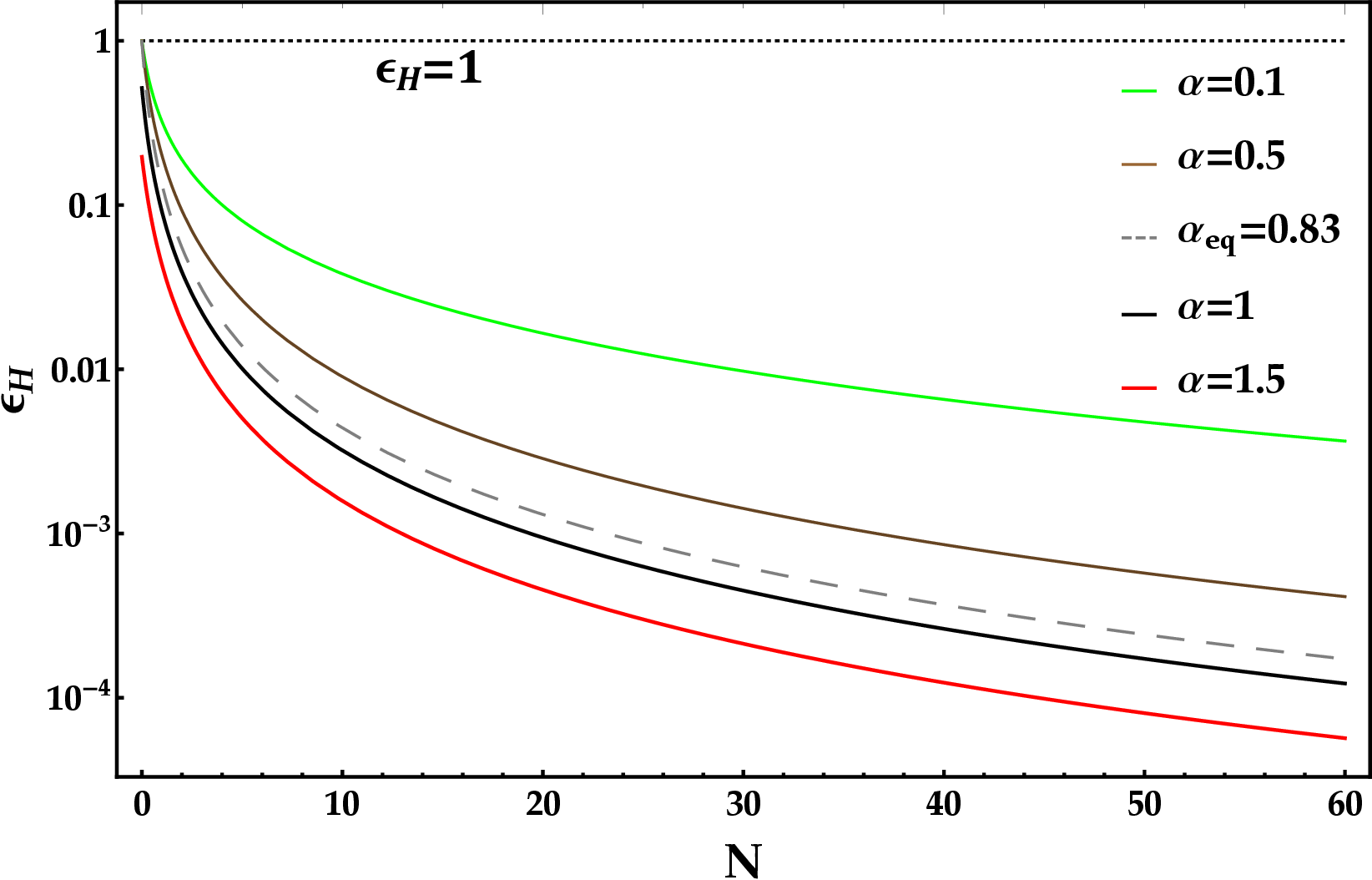}} \hspace{.1cm}
\centering
\subfigure[\label{etaN}]{\includegraphics[width=.48\textwidth]%
{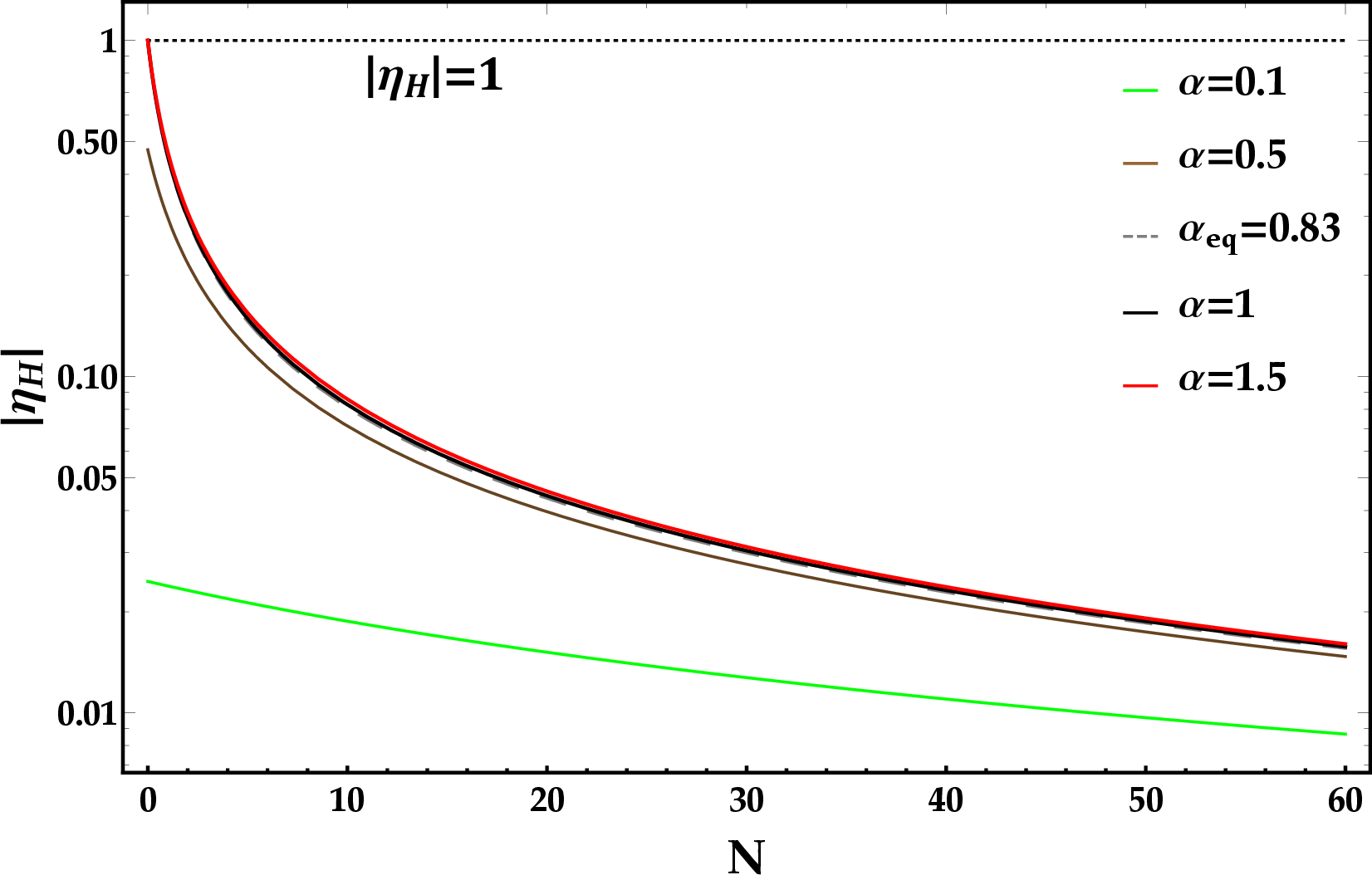}} \hspace{.1cm}
\end{minipage}
\caption{Evolutions of (a) the scalar field $\phi$, (b) the Hubble parameter $H$, (b) the first slow-roll parameter $\epsilon_{\rm H}$, and (d) the second slow-roll parameter $\eta_{\rm H}$ for different values of $\alpha$. The green, brown, black, and red lines represent $\alpha = 0.1, 0.5, 1$ and $1.5$, respectively. Also the dashed curve in each panel corresponds to $\alpha_{\rm eq}=0.83$. The end of inflation is set at $N=0$.}
\label{fig:back}
\end{figure}

Figure \ref{fig:back} illustrates evolutions of the scalar field, the Hubble parameter and the slow-roll parameters  (\ref{eq:epsilon})-(\ref{eq:eta}) in terms of the $e$-fold number $N$ for different values of $\alpha$.
Here, we utilize the $e$-fold definition ${\rm d}N=-H{\rm d}t$, and set the end of inflation at $N=0$. Figures \ref{phiN} and \ref{HN} show that (i) for a given $\alpha$, both the scalar field $\phi$ and the Hubble parameter $H$ decrease when the $e$-fold number $N$ decreases; (ii) for a given $N$, when the parameter $\alpha$ increases then $\phi$ and $H$ decrease.
 Figures \ref{eN} and \ref{etaN} illustrate that for $\alpha= \alpha_{\rm eq}\simeq 0.83$, we have $\epsilon_{\rm H}=|\eta_{\rm H}|=1$ at the end of inflation $N=0$. Besides, for  $\alpha>\alpha_{\rm eq}$ the condition of end of inflation is provided by $|\eta_{\rm H}|=1$ which occurs before  $\epsilon_{\rm H}=1$.

\begin{figure}[H]
\centering
\vspace{-0.2cm}
\scalebox{0.7}[0.7]{\includegraphics{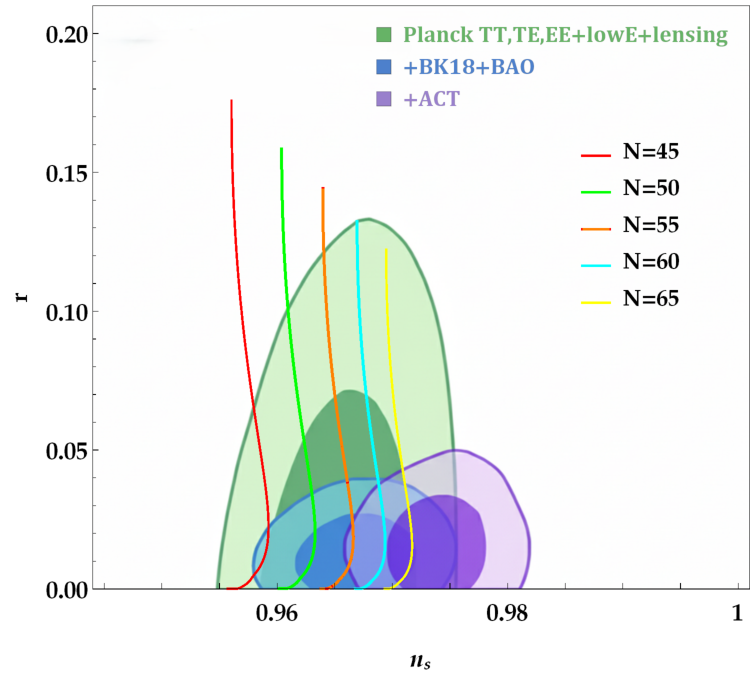}}
\vspace{-0.1cm}
\caption{Tensor-to-scalar ratio $r$ versus the scalar spectral index $n_{\rm s}$ for different $\alpha$ and various numbers of $e$-folds $N$.
The red, green, orange, blue, and yellow curves correspond to $N=45,~50,~55,~60$ and $65$, respectively.
The dark (light) green area represents the 68\% (95\%) CL from the Planck 2018 TT, TE, EE + LowE + Lensing data, while the dark (light) blue regions show the 68\% (95\%) CL from the combined Planck 2018 + BK18 + BAO dataset. The dark (light) purple area in the background corresponds to the 68\% (95\%) CL constraints from the ACT data (Planck 2018 + BK18 +ACT).
The  parameter $\alpha$ varies from 0 at the top to 5 at the bottom along each curve.}
\label{fig:r-ns}
\end{figure}

Note that the energy scale of inflation is defined by $V_0^{1/4}$ in Eq. (\ref{eq:MHI potential}). This energy scale can be obtained by fixing the scalar power spectrum in Eq. (\ref{eq:Ps-SR}) at the CMB scale ($k_\ast=0.05~{\rm Mpc^{-1}}$) and using Eqs. (\ref{eq:H2}) and (\ref{eq:epsilon}). The result reads
\begin{equation}\label{v0}
V_0^{1/4}=M_{\rm p}\left(\frac{12\pi^2\alpha^2{\cal P}_{\rm s}(k_*)\sinh(\alpha \phi_*) \tanh(\alpha \phi_*)}{[ \cosh(\alpha \phi_*)-1]^3}\right)^{1/4},
\end{equation}
where $\phi_*$ is the value of scalar field at the horizon exit. For the range of the parameter $\alpha\sim{\cal O}(0.1-1)$ considered in the present work, from Eq. (\ref{v0}) we estimate the energy scale of inflation as $V_0^{1/4}\sim{\cal O}(10^{-3})M_{\rm p}$.
To assess the compatibility of the mutated hilltop inflationary model with current observations, we constrain its parameters using the latest bounds on the scalar spectral index $n_{\rm s}$ and the tensor-to-scalar ratio $r$, given by Eq. (\ref{eq:nsr95}) and (\ref{eq:nsr95ACT}).
Figure \ref{fig:r-ns} illustrates the prediction of the mutated hilltop potential (\ref{eq:MHI potential}) in the $r-n_{\rm s}$ plane, in comparison with the observational constraints from the Planck 2018, BICEP/Keck 2018, and ACT datasets.
Each curve in the $r-n_{\rm s}$ panel corresponds to a specific $e$-fold number $N$, with the parameter $\alpha$ varying from 0 (at the top) to 5 (at the bottom)
along the curve.
From Fig. \ref{fig:r-ns}, the allowed range of $\alpha$ can be identified for each $N$.
For instance, for $N=50$ (green curve),  $\alpha$ is constrained to $\alpha \geq 0.206$ at the 95\% CL, and $0.291 \leq \alpha \leq 0.815$ at the $68\%$ CL, based on Planck 2018+ BK18 + BAO data (blue regions). The full results are summarized in Table \ref{tabTre} (Planck+BK18+BAO) and Table \ref{tabTreACT} (including ACT).
Applying the latest Planck 2018+ BK18 constraints (blue regions in Fig. \ref{fig:r-ns}) to the model predictions in the $r-n_{\rm s}$ plane yields lower bound on the inflationary duration of $N=44.4$  (95\% CL) and $N=48.1$ (68\% CL) (see Table \ref{tabTre}). This implies that for $N<44.4~(48.1)$, the model predictions lie outside the  $95\%$ (68\%) CL of the Planck+ BK18 data.
Including the ACT measurements (purple regions in Fig. \ref{fig:r-ns}) further tightens these limits and gives a lower bound of $N=54$  (95\% CL) (see Table \ref{tabTreACT}).
To further constrain the parameter space, we next examine the implications of the reheating and RD epochs on the model predictions in the following section.
\begin{table}[H]
  \centering
  \caption{Allowed ranges of the model parameter $\alpha$ related to the permitted $e$-fold numbers $N$ in the mutated hilltop inflationary model. The results are obtained by combining constraints from Planck 2018 and BICEP/Keck 2018 data ($r-n_{\rm s}$), the reheating parameter ($\omega_{\rm re}+N_{\rm re}+T_{\rm re}$), the RD era ($N_{\rm rd}$), and relic GWs.}
\resizebox{\textwidth}{!}{\fontsize{5}{6}\selectfont%
\begin{tabular}{c|c c|c c|c c}%
\thickhline
\multirow{2}{*}{$N$} &
\multicolumn{2}{c|} {$ (r-n_{\rm s}) $}&
\multicolumn{2}{c|} {\makecell[c]{$(r-n_{\rm s}) $ \\ $+ \omega_{\rm re}+N_{\rm re}+T_{\rm re}$ \\ $+N_{\rm rd} $}} &
\multicolumn{2}{c} {\makecell[c]{$ (r-n_{\rm s})  $ \\ $+ \omega_{\rm re}+N_{\rm re}+T_{\rm re}$ \\ $+N_{\rm rd}$ \\ $+ {\rm GWs}$}} \\[0.5ex] \cline{2-7}
 &
$\alpha ~(95 \% $ CL)  & $\alpha ~(68 \% $ CL) &
$\alpha ~(95 \% $ CL) & $\alpha ~(68 \% $ CL) &
$\alpha ~(95 \% $ CL) & $\alpha ~(68 \% $ CL) \\[0.5ex] \thickhline
$44.4$ &
$ [0.372,0.642] $ &
$ - $ &
$ [0.372,0.642] $ &
$  -  $  &
$  -  $  &
$  -  $ \\[0.5ex] 
$45$ &
$ [0.326,0.758] $ &
$ - $ &
$  [0.326,0.758]  $ &
$  -  $   &
$  -  $  &
$  -  $ \\[0.5ex] 
$46$ &
$ [0.282,1.179] $ &
$ - $ &
$ [0.282,1.179]  $  &
$  -  $ &
$ [0.779,0.890] $ &
$  -  $ \\[0.5ex] 
$48.1$ &
$ \alpha \geq 0.233 $ &
$ [0.410,0.490] $ &
$ [0.233,1.485] $  &
$ [0.410,0.490] $ &
$ [0.362,0.855] $ &
$ [0.410,0.490] $ \\[0.5ex] 
$ 50 $  &
$ \alpha \geq 0.206 $ &
$ [0.291,0.815] $ &
$ [0.206,1.485] $ &
$ [0.291,0.815] $ &
$ [0.206,0.818] $ &
$ [0.291,0.815] $ \\[0.5ex] 
$55$ &
$ \alpha \geq 0.166 $ &
$ \alpha \geq 0.222 $ &
$ [ 0.166,1.485] $ &
$ [ 0.222,1.485] $ &
$ [0.166,0.721] $ &
$ [0.222,0.721] $ \\[0.5ex]
$56$ &
$ \alpha \geq 0.161 $ &
$ \alpha \geq 0.217 $ &
$[ 0.161,0.522] $ &
$ [ 0.217,0.522] $ &
$ [0.161,0.522] $ &
$  [0.217,0.522] $ \\[0.5ex] 
$57$ &
$ \alpha \geq 0.157 $ &
$ \alpha \geq 0.212 $ &
$ - $ &
$ - $ &
$ - $ &
$ - $ \\[0.5ex]  
$60$ &
$ \alpha \geq 0.147 $ &
$ \alpha \geq 0.204 $ &
$ - $ &
$ - $ &
$ - $ &
$  -  $ \\[0.5ex] 
$65$ &
$ \alpha \geq 0.138 $ &
$ [0.244,0.695] $ &
$ - $ &
$ - $ &
$ - $ &
$ - $ \\[0.5ex] \thickhline
\end{tabular}}
 \label{tabTre}
\end{table}
\begin{table}[H]
  \centering
  \caption{Allowed ranges of the model parameter $\alpha$ related to the permitted $e$-fold numbers $N$ in the mutated hilltop inflationary model obtained by combining  constraints from the Planck 2018, BICEP/Keck 2018, and ACT data ($r-n_{\rm s}$), reheating parameters ($\omega_{\rm re}+N_{\rm re}+T_{\rm re}$), the RD era ($N_{\rm rd}$), and relic GWs.}
\resizebox{\textwidth}{!}{\fontsize{5}{6}\selectfont%
\begin{tabular}{c|c c|c c|c c}%
\thickhline
\multirow{2}{*}{$N$} &
\multicolumn{2}{c|} {$ (r-n_{\rm s}) $}&
\multicolumn{2}{c|} {\makecell[c]{$(r-n_{\rm s}) $ \\ $+ \omega_{\rm re}+N_{\rm re}+T_{\rm re}$ \\ $+N_{\rm rd} $}} &
\multicolumn{2}{c} {\makecell[c]{$ (r-n_{\rm s})  $ \\ $+ \omega_{\rm re}+N_{\rm re}+T_{\rm re}$ \\ $+N_{\rm rd}$ \\ $+ {\rm GWs}$}} \\[0.5ex] \cline{2-7}
 &
$\alpha ~(95 \% $ CL)  & $\alpha ~(68 \% $ CL) &
$\alpha ~(95 \% $ CL) & $\alpha ~(68 \% $ CL) &
$\alpha ~(95 \% $ CL) & $\alpha ~(68 \% $ CL) \\[0.5ex] \thickhline
$ 54 $  &
$ [0.29,0.38] $ &
$ - $ &
$ [0.29,0.38] $ &
$ - $ &
$ [0.29,0.38] $ &
$ - $ \\[0.5ex] 
$55$ &
$ [ 0.245,0.50] $ &
$ - $ &
$ [ 0.245,0.50] $ &
$ - $ &
$ [ 0.245,0.50] $ &
$ - $ \\[0.5ex]
$56$ &
$ [ 0.216,0.62] $ &
$ - $ &
$ [ 0.216,0.62] $ &
$ - $ &
$ [ 0.216,0.62] $ &
$  - $ \\[0.5ex] 
$57$ &
$ [ 0.194,0.78] $ &
$ - $ &
$ - $ &
$ - $ &
$ - $ &
$ - $ \\[0.5ex]  
$60$ &
$ \alpha \geq 0.155 $ &
$ - $ &
$ - $ &
$ - $ &
$ - $ &
$  -  $ \\[0.5ex] 
$65$ &
$ \alpha \geq 0.125 $ &
$ [0.17,0.85] $ &
$ - $ &
$ - $ &
$ - $ &
$ - $ \\[0.5ex] \thickhline
\end{tabular}}
 \label{tabTreACT}
\end{table}

\section{Reheating implications} \label{sec3}
Inflation is followed by a critical phase known as reheating.
During this phase, the inflaton field begins to oscillate around the potential minimum and subsequently decays into standard model particles which leads to the thermalization of the Universe \cite{Kofman:1994,Martin:2010,Unnikrishnan:2012,LiangDai:2014,Cook:2015,Gialamas:2020,German:2023,sahni:1990}. Thereafter, the Universe becomes hot enough to initiate the radiation dominated era.
The duration of the reheating epoch $N_{\rm re}$ is influenced by the rate of inflaton decay.
Meanwhile, the reheating temperature $T_{\rm re}$ is sensitive to $N_{\rm re}$ and, consequently, depends on the inflaton decay rate as well. It is known that a longer reheating phase corresponds to a lower reheating temperature. Both  $N_{\rm re}$ and $T_{\rm re}$  depend on the equation of state parameter $\omega_{\rm re}$ through the following model dependent relations \cite{German:2023}
\begin{eqnarray}
&N_{\rm re}=\left(\frac{-4}{1-3 \omega _{\rm re}}\right)\left[ N+\frac{1}{3} \ln \left( \frac{11 g_{\rm re}^{\rm s}}{43} \right)+\frac{1}{4}\ln \left( \frac{30}{\pi ^2g_{\rm re}} \right)+\ln \left( \frac{k_\ast}{a_0 T_0} \right)+\ln \left( \frac{\rho_{\rm e}^{\frac{1}{4}}}{H_{\ast}} \right)\right], \label{eq:Nre1}\\[0.3cm]
&T_{\rm re} =\Big( \frac{30 \rho_{\rm e}}{\pi^2 g_{\rm re}} \Big)^{\frac{1}{4}} e^{-\frac{3}{4}(1+\omega_{\rm re})N_{\rm re}}.
\label{eq:Tre}
\end{eqnarray}
Here, $g_{\rm re}$ and $g_{\rm re}^{\rm s}$ represent the relativistic degrees of freedom. For the temperature range $10~{\rm MeV} \leq T_{\rm re} < 200~{\rm GeV}$, both $g_{\rm re}$ and $g_{\rm re}^{\rm s}$ are generally temperature dependent and vary within the range $10.75 \leq g_{\rm re} = g_{\rm re}^{\rm s} < 106.75$. However, the small effect of their temperature variations on $N_{\rm re}$ can be neglected \cite{Mishra:2021}. Therefore, we set $g_{\rm re} = g_{\rm re}^{\rm s} = 106.75$ for $T_{\rm re} \geq 200~{\rm GeV}$.
Additionally, $a_0=1$ is the current scale factor, $T_0=2.725~{\rm K}$ is the present temperature of the CMB, and $k_\ast=0.05~{\rm Mpc^{-1}}$ signifies the comoving wavenumber at the pivot scale.
Furthermore, $\rho_{\rm e}=\rho_{\phi_{\rm e}}$ represents the energy density of inflaton at the end of inflation. By using Eqs. (\ref{eq:rho}) and (\ref{eq:Lp}) and setting $\omega_\phi\equiv p_{\phi}/\rho_{\phi}=-1/3$ at the end of inflation, one can  obtain
\begin{equation}\label{rhoe}
\rho_{\rm e}=\frac{3}{2} V_{\rm e},
\end{equation}
where $V_{\rm e}=V(\phi_{\rm e})$ is the potential at the end of inflation.

\begin{figure}[H]
\centering
\vspace{-0.2cm}
\scalebox{0.5}[0.5]{\includegraphics{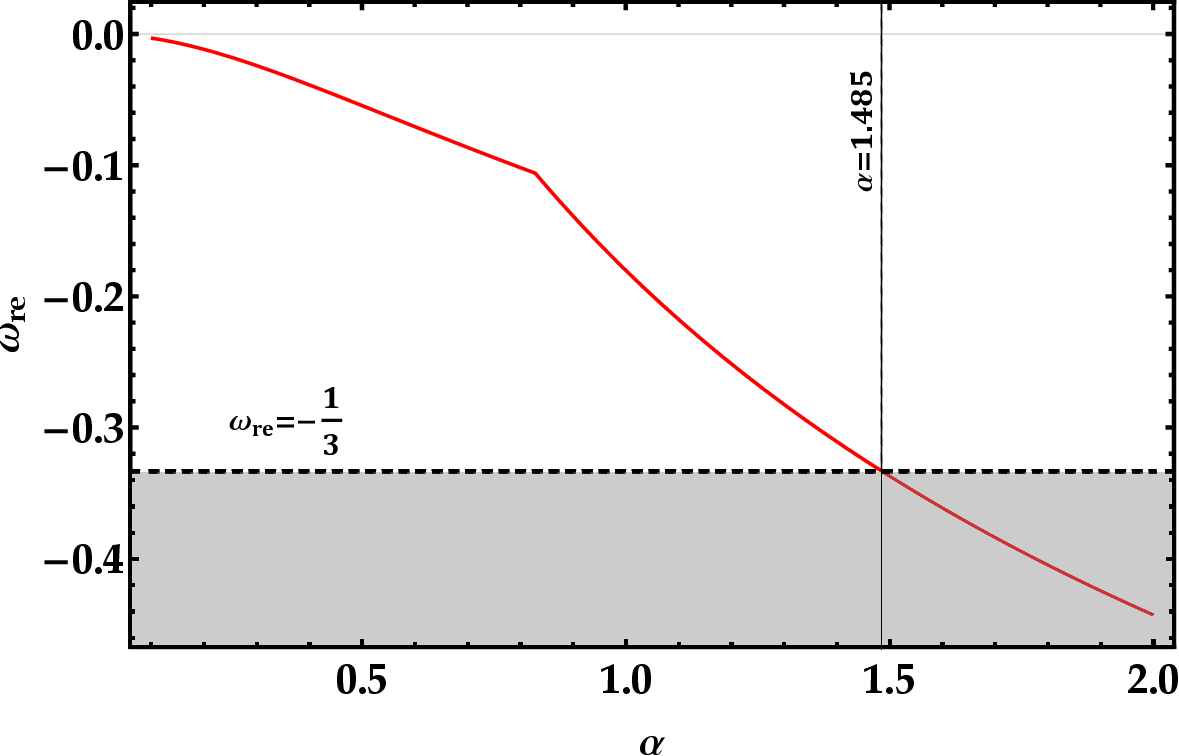}}
\vspace{-0.1cm}
\caption{Variation of the equation of state parameter, $\omega_{\rm re}$, versus the model parameter, $\alpha$.
The lower bound of $\omega_{\rm re}$ at $-1/3$ is shown as a horizontal dashed line, while the upper bound of $\alpha$ at $1.485$ is denoted by a vertical line.
The shaded region represents disallowed values of $\omega_{\rm re}$.}
\label{fig:wre}
\end{figure}

\begin{figure}[H]
\begin{minipage}[b]{0.9\textwidth}
\centering
\subfigure[\label{Nre_ns} ]{ \includegraphics[width=0.40\textwidth]%
{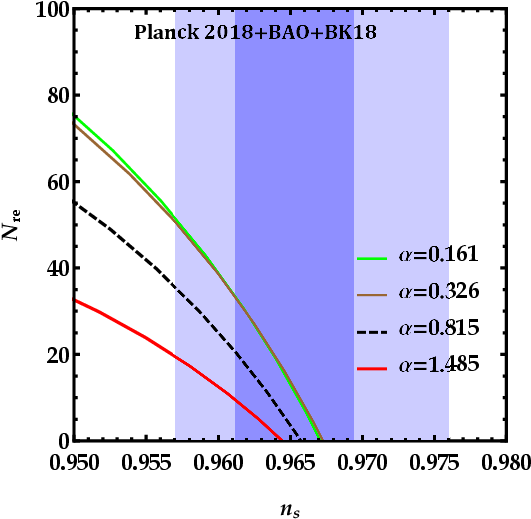}}
\subfigure[\label{Nre_nsACT} ]{ \includegraphics[width=0.40\textwidth]%
{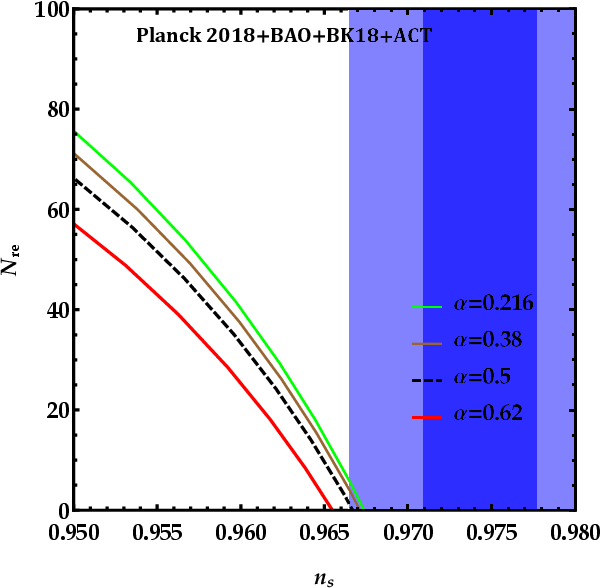}}\\
\subfigure[\label{Nre_a}]{ \includegraphics[width=0.40\textwidth]%
{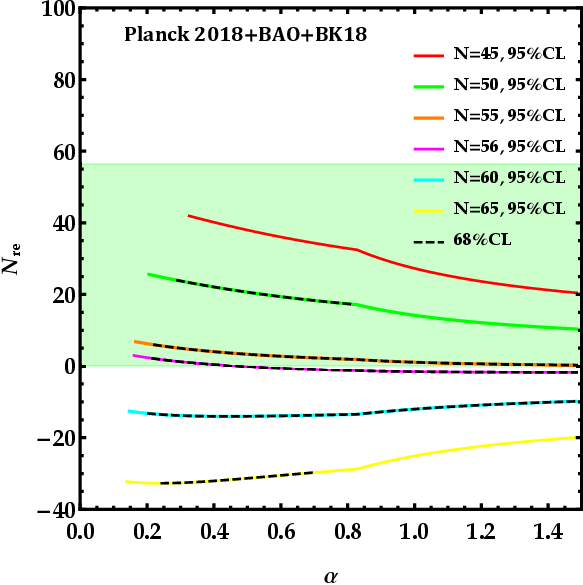}}
\subfigure[\label{Nre_aACT}]{ \includegraphics[width=0.40\textwidth]%
{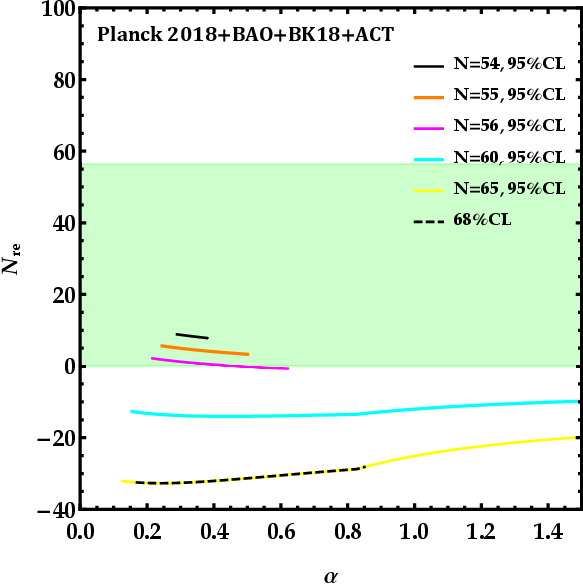}}\\
\subfigure[\label{Nre_wre} ]{ \includegraphics[width=0.40\textwidth]%
{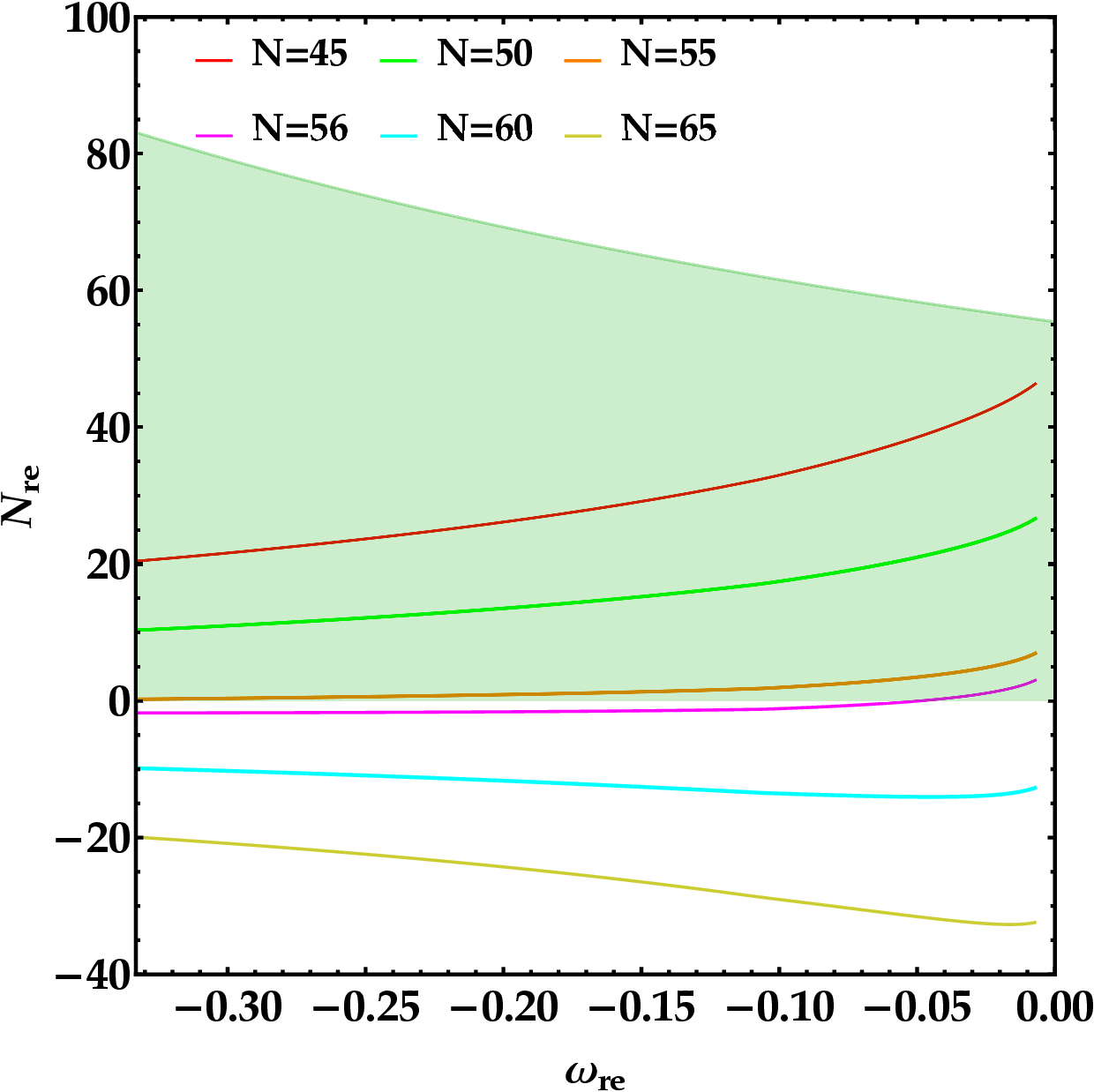}}
\end{minipage}
\caption{Variations of $N_{\rm re}$ vs (a, b) $n_{\rm s}$, (c, d)  $\alpha$, and (e) $\omega_{\rm re}$.
The dark (light) blue regions in panels (a) and (b) denote the 68\% (95\%) CL from the Planck 2018  + BK18 + BAO data and from the combined  Planck 2018  + BK18 + BAO + ACT dataset, respectively.
The light green shaded regions in panels (c)-(d) indicate the model independent bounds on  $N_{\rm re}$ for $\omega_{\rm re}=0$ while the similar region in panel (e) corresponds to the model independent bound on $N_{\rm re}$ for $-1/3\leq\omega_{\rm re}<0$.}
\label{fig:Nre}
\end{figure}

\begin{figure}[H]
\begin{minipage}[b]{1\textwidth}
\centering
\subfigure[\label{Tre_ns} ]{ \includegraphics[width=0.48\textwidth]%
{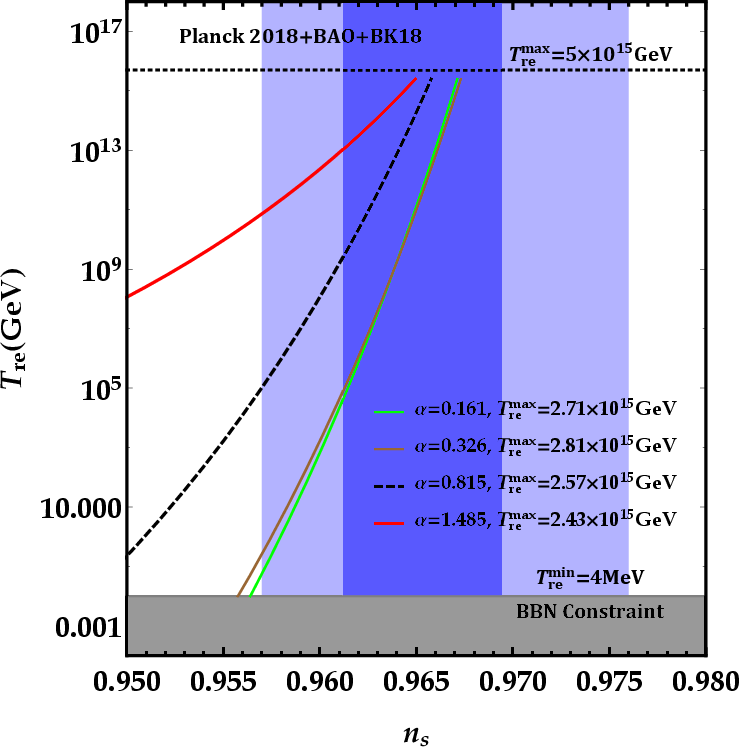}}\hspace{.1cm}
\subfigure[\label{Tre_nsACT} ]{ \includegraphics[width=0.48\textwidth]%
{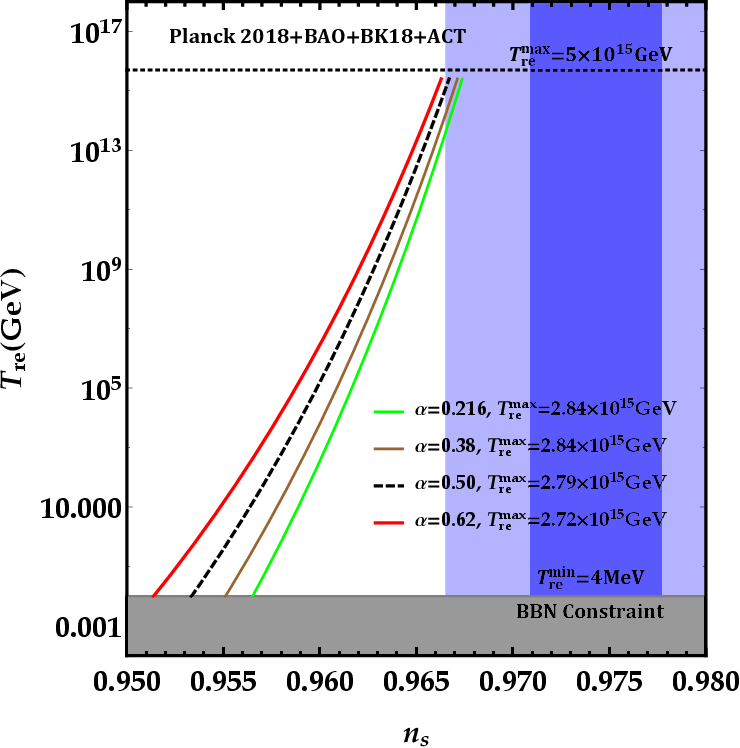}}\hspace{.1cm}\\
\subfigure[\label{Tre_a}]{ \includegraphics[width=.48\textwidth]%
{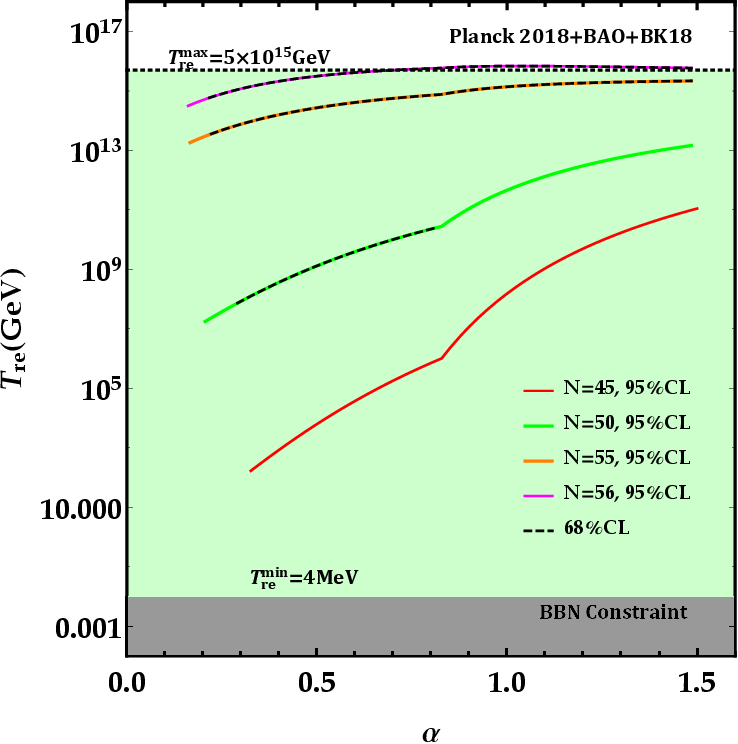}}
\subfigure[\label{Tre_aACT}]{ \includegraphics[width=.48\textwidth]%
{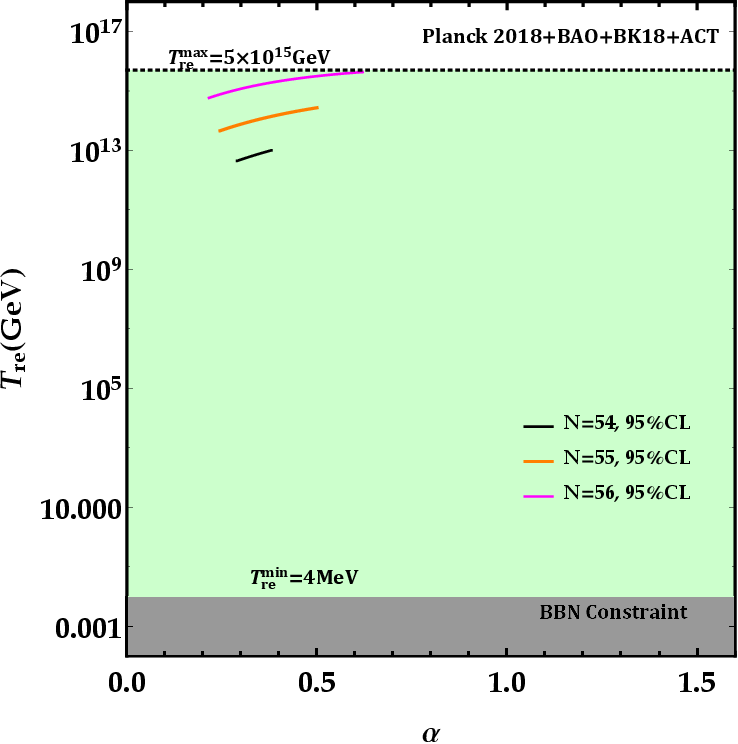}}
\end{minipage}
\caption{Variations of the reheating temperature $T_{\rm re}$ versus (a, b) the scalar spectral index $n_{\rm s}$, and (c, d) the model parameter $\alpha$.
The shaded blue regions in panels (a) and (b) are the same as in Fig. \ref{Nre_ns}.
The light green region in panels (c) and (d) indicate the allowed range of $T_{\rm re}$ from the model independent constraint in Eq. (\ref{eq:Tre MIB}).
The horizontal black dashed line represents the maximum allowed reheating temperature $T_{\rm re}^{\rm max}=5\times 10^{15}~\rm GeV$, while the bottom hatched area corresponds to the BBN constraint.
}\label{fig:Tre1}
\end{figure}

Moreover, $H_*$ in Eq. (\ref{eq:Nre1}) denotes the Hubble parameter at the pivot scale and  can be estimated using Eq. (\ref{eq:Ps-SR}) as follows
\begin{equation}\label{eq:Hi}
H_{\ast}=2\pi M_{\rm p}\sqrt{ 2 {\cal P}_{\rm s}(k_\ast ) \epsilon _{\rm H}(\phi_*) },
\end{equation}
where ${\cal P}_{\rm s}(k_\ast ) = 2.1\times 10^{-9}$ is the amplitude of scalar power spectrum at the CMB pivot scale.
 As for the equation of state parameter $\omega_{\rm re}$ appeared in Eq. (\ref{eq:Nre1}), the range of $-1/3 \leq \omega_{\rm re} \leq 1$ is considered during the reheating epoch \cite{Cook:2015}.
The lower bound $\omega_{\rm re}=-1/3$ arises from the end of inflation condition, i.e., $\epsilon_{\rm H}=1$,
while the upper bound $\omega_{\rm re}=1$ represents the most conservative limit derived from causality considerations. Given the broad range of possible values for $\omega_{\rm re}$, it is challenging to  accurately constraining the model parameter space. So, the validity of the model predictions could be compromised.
In order to enhance the reliability of our model, we seek to establish a correlation between $\omega_{\rm re}$ and the free parameter $\alpha$.
As demonstrated in \cite{Unnikrishnan:2012}, this correlation can be estimated as follows
\begin{equation}\label{eq:wre}
1+\left \langle \omega _{\phi} \right \rangle= 2 \left [ \int_{0}^{\phi _{\rm m}}{\mathrm{d}\phi \left ( 1-\frac{V\left (\phi  \right ) }{V\left (\phi_{\rm m}  \right )} \right ) }^{\frac{1}{2 }}\right ]\left [\int_{0}^{\phi _{\rm m}}{\mathrm{d}\phi \left ( 1-\frac{V\left (\phi  \right ) }{V\left (\phi_{\rm m}  \right )} \right ) }^{-\frac{1}{2 }}  \right ]^{-1} ,
\end{equation}
where $\langle \omega_{\phi} \rangle$ denotes the average value of the equation of state parameter over an oscillation cycle, which is equivalent to the reheating equation of state parameter, i.e., $\langle \omega_{\phi} \rangle=\omega_{\rm re}$.
Additionally, the parameter $\phi_{\rm m}$ represents the maximum value of the scalar field during that cycle.
We integrate Eq. (\ref{eq:wre}), numerically, and present the result in Fig. \ref{fig:wre}.
The horizontal dashed line in this figure indicates the lower bound of the reheating equation of state parameter $\omega_{\rm re}=-1/3$.
Our calculations from Eq. (\ref{eq:wre}) reveal that this lower bound corresponds to $\alpha=1.485$, which is considered as the maximum value of $\alpha$.
Furthermore, it can be seen from Fig. \ref{fig:wre} that due to the positivity of $\alpha$, the maximum value of $\omega_{\rm re}$ approaches zero. Hence, the equation of state parameter in our model is constrained to the range  $-1/3\leq\omega_{\rm re}<0$ which results in an upper bound on the $\alpha$ parameter as $\alpha\leq1.485$. Note that the shaded regions in gray, are excluded due to $\omega_{\rm re}<-1/3$.
\vspace{.5cm}
Now,  utilizing this model dependent approach in Eqs. (\ref{eq:Nre1})-(\ref{eq:wre}), we can compute the length and temperature of the reheating period as functions of $\alpha$.
Figures \ref{Nre_ns}-\ref{Nre_nsACT} and \ref{Tre_ns}-\ref{Tre_nsACT} illustrate the variations of $N_{\rm re}$ and $T_{\rm re}$ with respect to $n_{\rm s}$ for different values of $\alpha$. In Figs. \ref{Nre_ns} and \ref{Tre_ns}, the background regions correspond to the Planck 2018 + BK18 data, whereas in Figs. \ref{Nre_nsACT} and \ref{Tre_nsACT}, they are obtained using the combined Planck 2018 + BK18 + ACT dataset.
Note that the length of reheating period cannot be negative, i.e. $N_{\rm re}\geq 0$.
Furthermore, as demonstrated in \cite{German:2023}, a model independent correlation exists between the reheating parameters, which is given by
\begin{equation}\label{eq:Nre MIB}
N_{\rm re}^{\rm MIB}=\ln \left [  \frac{\Big(\frac{43}{11 g_{\rm re}^{\rm s}}\Big)^{\frac{2}{3}}\pi \sqrt{{\cal P}_{\rm s}(k_\ast) r}~T_{0}^{2}}{\sqrt{2}H_0\sqrt{\Omega _{\rm r_0}} ~T_{\rm re}^2} \right ]^\frac{2}{3(1+\omega _{\rm re})},
\end{equation}
where the superscript "$\rm MIB$" signifies that this is a model independent bound. Also $g_{\rm re}^{\rm s}$ is the number of relativistic degrees of freedom in entropy at the reheating epoch. The value of $g_{\rm re}^{\rm s}$ is typically considered to be 10.75 for low temperatures around $ T_{\rm re} = 4~ \rm MeV$, and 106.75 for $ T_{\rm re}\geqslant 200~ \rm GeV $ \cite{German:2023}. In addition, $\Omega_{\rm r_0}=2.47\times10^{-5}h^{-2}$, $T_0=2.725~\rm K$ and $H_0=67.4~\rm Km~ s^{-1}~ Mpc^{-1}$ are the current values of the radiation density parameter, the CMB temperature and the Hubble parameter, respectively \cite{planck:2020}. Furthermore, in Eq. (\ref{eq:Nre MIB}) we set the tensor-to-scalar ratio as $r=r_{\rm max}=0.036$ \cite{bk18}.

In the model independent approach \cite{German:2023}, the constraint on the reheating temperature reads
\begin{equation}\label{eq:Tre MIB}
4~{\rm MeV} \leq T_{\rm re}^{\rm MIB}\leq 5\times 10^{15} {~\rm GeV},
\end{equation}
where the lower bound comes from the Big Bang Nucleosynthesis (BBN) constraint \cite{Kawasaki:1999,Kawasaki:2000,Hasegawa}. Also the upper limit is obtained by setting $N_{\rm re}=0$  and $g_{\rm re}^{\rm s}=106.75$ in Eq. (\ref{eq:Nre MIB}), which is interestingly independent of $\omega_{\rm re}$.

Now, by substituting $T_{\rm re}^{\rm min}=4~ \rm MeV$ and  $g_{\rm re}^{\rm s}=10.75$ in Eq. (\ref{eq:Nre MIB}), we can derive an upper bound for $N_{\rm re}$ as a function of $\omega_{\rm re}$ which is plotted as a light green region in Fig. \ref{Nre_wre}. In other words, this region is shaded between two curves including $N_{\rm re}=0$ (for
$T_{\rm re}^{\rm max}=5 \times 10^{15}{~\rm GeV}$) and $N_{\rm re}$ as a function of $\omega_{\rm re}$ (for $T_{\rm re}^{\rm min}=4~ \rm MeV$). Furthermore,  Fig. \ref{Nre_wre} illustrates variation of $N_{\rm re}$ versus $\omega_{\rm re}$ for different values of the inflationary durations $N$ according to Table \ref{tabTre}. It is inferred from this figure that (i) $N=60$ (blue curve) and $N=65$ (yellow curve) are ruled out due to their relevant negative duration of reheating $N_{\rm re}<0$; (ii) $N=56$ (pink curve) is allowed for the duration of inflation only within $-0.058 \leq \omega_{\rm re}< -0.008 $; (iii) the allowed range for the duration of inflation constrained to $44.4\leq N\leq 56$ (95\% CL) and
$48.1\leq N\leq 56$ (68\% CL)  (see the third column of Table \ref{tabTre}); and
(iv) $\omega_{\rm re}=0$ imposes a severe constraint on the reheating duration as  $0\leq N_{\rm re}\leq 55.4$ compared to $0\leq N_{\rm re}\leq 83.1$ for $\omega_{\rm re}=-1/3$ (see the light green region corresponding to the model independent bound in Fig. \ref{Nre_wre}).
%
Therefore, we adopt the interval  $0\leq N_{\rm re}\leq 55.4$ as the model independent bound in Figs. \ref{Nre_a}-\ref{Nre_aACT}.

Figures \ref{Nre_a} and \ref{Nre_aACT} show the behaviour of $N_{\rm re}$ as a function of $\alpha$ for different numbers of $N$, based on Tables \ref{tabTre} (Planck 2018 + BK18) and \ref{tabTreACT} (Planck 2018 + BK18 + ACT). In these figures,
the solid (dashed) lines represent the allowed range of $\alpha$ at the $95\%$ ($68\%$) CL for different $N$.
By analyzing these figures, the specific values of $\alpha$  consistent with the $N_{\rm re}$ constraints of the present model can be identified. It is evident that for  $N=56$, the permissible range of $\alpha$ is $0.161 \leq \alpha \leq 0.522$ at the 95\% CL and $0.217\leq \alpha \leq 0.522$ at the 68\% CL (see the third column of Table \ref{tabTre}). When the ACT data are included, the range becomes $0.216 \leq \alpha \leq 0.62$ at the 95\% CL (see the third column of Table \ref{tabTreACT}).

In what follows, by setting $N_{\rm re}=0$ in the model dependent expression (\ref{eq:Tre}), we calculate the maximum reheating temperature for the various values of $\alpha$. The results are depicted in the legends of Figs. \ref{Tre_ns} and \ref{Tre_nsACT}.
The horizontal dashed line in panels of Fig. \ref{fig:Tre1} represents the model independent upper limit of the reheating temperature, $T_{\rm re}^{\rm max}=5 \times 10^{15}{~\rm GeV}$, while the curves illustrate the model predictions.
Figures \ref{Tre_a} and \ref{Tre_aACT} show the behaviour of the reheating temperature $T_{\rm re}$ as a function of the parameter $\alpha$ for different numbers of $N$.
These figures demonstrate that the allowed range of $\alpha$ for $N=56$ (pink curve) is consistent with the results previously obtained from the condition  $N_{\rm re} \geq 0$ (see the third columns of Tables \ref{tabTre} and \ref{tabTreACT}). It is obvious that, within the specific range of  $\alpha$ given in  Tables \ref{tabTre} and \ref{tabTreACT}, the reheating temperature $T_{\rm re}$ remains below the upper bound $T_{\rm re}^{\rm max}=5 \times 10^{15}{~\rm GeV}$. Consequently, applying the model independent reheating constraints, Eqs. (\ref{eq:Nre MIB}) and (\ref{eq:Tre MIB}) imposes an upper limit on the inflationary duration in this model as $N\leq 56$ (see the third columns of Tables \ref{tabTre} and \ref{tabTreACT}).
\subsection{Radiation dominated consideration}
\begin{figure}[H]
\begin{minipage}[b]{1\textwidth}
\centering
\subfigure[\label{Nrd_ns} ]{ \includegraphics[width=0.33\textwidth]%
{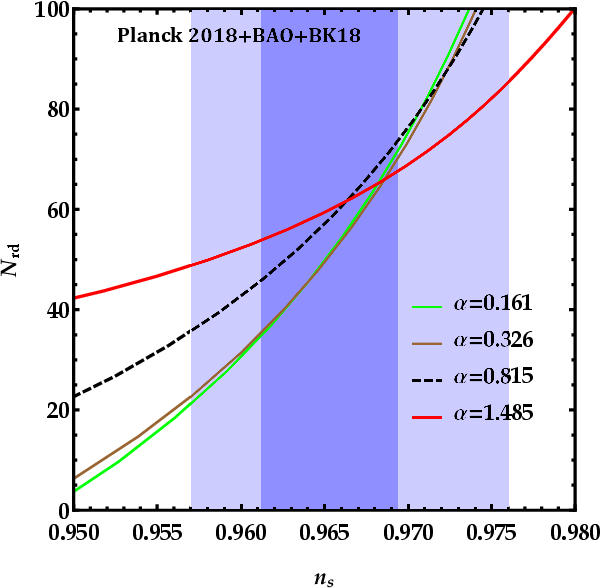}}
\subfigure[\label{Nrd_nsACT} ]{ \includegraphics[width=0.33\textwidth]%
{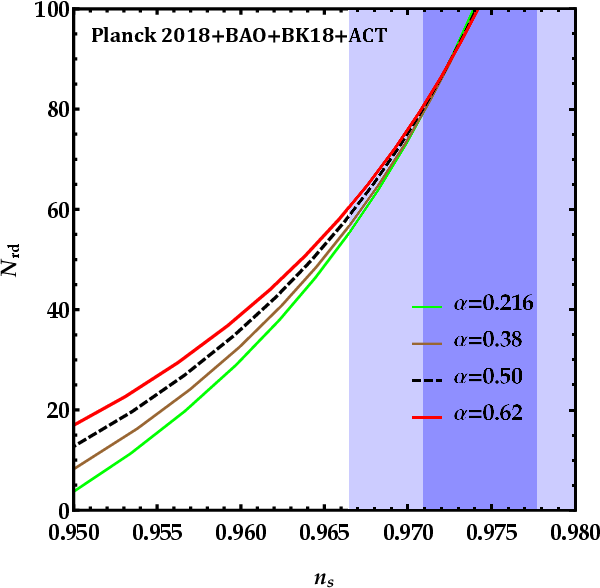}}\\
\subfigure[\label{Nrd_a}]{ \includegraphics[width=0.33\textwidth]%
{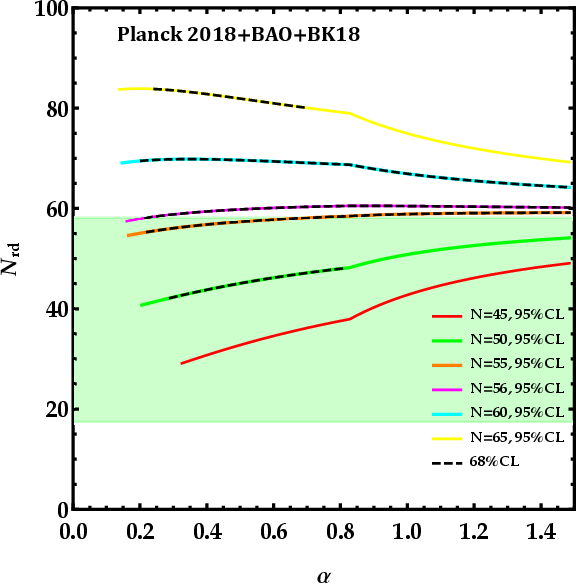}}
\subfigure[\label{Nrd_aACT}]{ \includegraphics[width=0.33\textwidth]%
{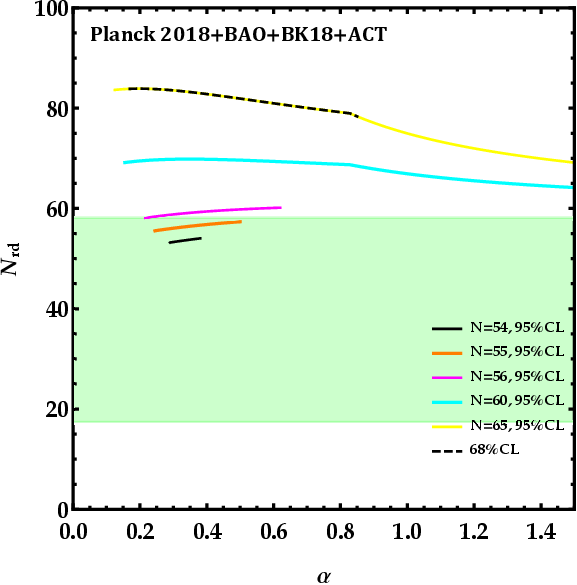}}\\
\subfigure[\label{Nrd_wre} ]{ \includegraphics[width=0.33\textwidth]%
{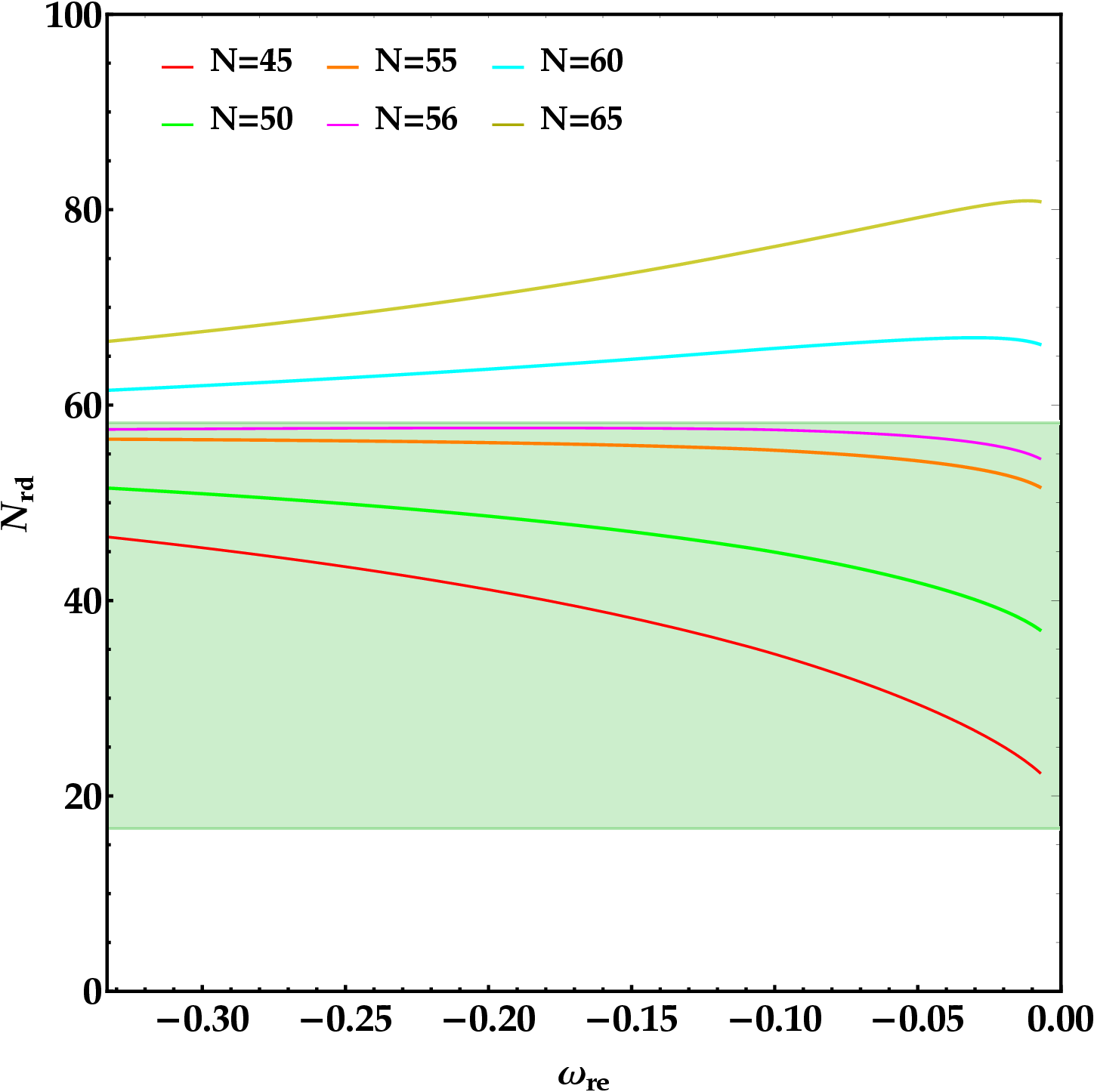}}

\end{minipage}
\caption{ Variations of the RD duration $N_{\rm rd}$ with respect to (a, b) $n_{\rm s}$, (c, d)  $\alpha$, and (e) $\omega_{\rm re}$.
The shaded regions in panels  (a) and (b) are the same as in Fig. \ref{Nre_ns}.
The light green shaded regions in graphs (c), (d) and (e) indicate the model independent bound from Eq. (\ref{Nrd}).}
\label{fig:Nrd}
\end{figure}

As previously mentioned, after thawing the Universe  during the reheating epoch, the RD era is initiated. It has been shown that the reheating parameters can influence the duration of the RD era, $N_{\rm rd}$, as follows \cite{German:2023}
\begin{equation}
N_{\rm rd} =  -\frac{3(1+\omega _{\rm re})}{4}N_{\rm re}+\frac{1}{4} \ln \left ( \frac{30}{g_{\rm re}\pi ^{2}} \right )+\frac{1}{3}\ln \left( \frac{11g_{\rm re}^{\rm s}}{43} \right )+\ln \left ( \frac{a_{\rm eq}\rho _{\rm e}^{\frac{1}{4}}}{a_0 T_0} \right ),
\label{eq:Nrd MIB}
\end{equation}
where $a_{\rm eq}=2.94 \times 10^{-4}$ is the scale factor at the time of equality radiation-matter. Replacing Eqs. (\ref{eq:Nre1}) and (\ref{eq:wre}) into (\ref{eq:Nrd MIB}), one can numerically determine the length of the RD period as a function of $\alpha$.
Figures \ref{Nrd_ns} and \ref{Nrd_nsACT} illustrate variation of $N_{\rm  rd}$ with respect to $n_{\rm s}$ for different values of $\alpha$.

Additionally, as shown in \cite{German:2023}, a model independent correlation exists between the reheating temperature and the length of the RD phase $N_{\rm rd}$, which is given by
\begin{equation}
N_{\rm rd}^{\rm MIB} =\ln \left [ \frac{a_{\rm eq}T_{\rm re}}{\Big ( \frac{43}{11 g_{\rm re}^{\rm s}} \Big )^\frac{1}{3} T_0} \right ].
\label{eq:Nrd}
\end{equation}
Substituting the model independent bound on the reheating temperature (\ref{eq:Tre MIB}) into Eq. (\ref{eq:Nrd}), one can determine the allowed range for the length of RD era as
\begin{equation}\label{Nrd}
16.7\leq N_{\rm rd}^{\rm MIB}\leq 58.2,
\end{equation}
where the value of $g_{\rm re}^{\rm s}$ is considered to be 10.75 and 106.75 for $ T_{\rm re}^{\rm min}$ and $ T_{\rm re}^{\rm max}$, respectively.
In Figs. \ref{Nrd_a}-\ref{Nrd_wre}, the light green regions represent the parameter space constrained by the bound (\ref{Nrd}).
As it is evident from these figures, incorporating the implications of the RD era leads to no modification of the previously estimated bounds on $N$ and $\alpha$ deduced from the combined constraints $(r-n_{\rm s})+\omega_{\rm re}+N_{\rm re}+T_{\rm re}$ (see again the third columns of Tables \ref{tabTre} and \ref{tabTreACT}).

Notably, the findings of Ref. \cite{German:2023} suggest a universal upper bound of 56 $e$-folds for observable inflation, with a maximum uncertainty of one $e$-fold. This result is in excellent agreement with the predictions of the present model, where a combination of constraints from the CMB, reheating, and the RD era confines the inflationary duration to $44.4\leq N\leq 56$ (95\% CL) and $48.1\leq N\leq 56$ (68\% CL) based on the Planck + BK18 + BAO data, and $54\leq N\leq 56$ (95\% CL) when the ACT measurements are included (see the third columns of Tables \ref{tabTre} and \ref{tabTreACT}).

To further constrain the free parameter $\alpha$ of the present model, we will investigate the possibility of generating GWs in the following section. Our aim is to identify the specific range of $\alpha$ values that could lead to the production of GWs detectable by GW observatories.


\section{The relic gravitational waves} \label{sec4}
A decisive prediction of inflation theory is the generation of relic gravitational waves.
During inflation, tensor perturbations exit the horizon and subsequently re-enter after the inflationary epoch.
These re-entering tensor perturbations can give rise to the propagation of GWs \cite{Mishra:2021,allen88,sahni:1990,sami2002,dany_18,dany_19,Bernal:2019lpc}.
It is noteworthy that relic GWs offer a unique probe into the early Universe. Due to their minimal interaction with matter and radiation, they propagate undisturbed and carry valuable information about the Universe primordial conditions.
The current energy density spectrum of these primordial GWs in the reheating epoch is related to  its counterpart in the RD era as follows \cite{Mishra:2021}
\begin{equation}
\Omega _{\rm GW_0}^{(\rm re)}(f)  = \Omega _{\rm GW_0}^{(\rm RD)}(f) \left (\frac{f}{f_{\rm re}}\right )^{2\left (\frac{\omega_{\rm re}-1/3}{\omega_{\rm re}+1/3}\right )},~~~ f_{\rm re} < f \leq f_{\rm e}.
\label{eq:GW_spectrum_2b}
\end{equation}
Here, $\Omega _{\rm GW_0}^{(\rm RD)}$ is the current energy density spectrum of GWs in the RD era and is given by
\begin{equation}
\Omega _{\rm GW_0}^{(\rm RD)}(f)=\left ( \frac{1}{24} \right )r~{\cal P}_{\rm s} (k_{\ast })\left ( \frac{f}{f_{\ast }} \right )^{n_{{\rm t}}}\Omega _{\rm r_0},~~~f_{\rm eq}< f\leq f_{\rm re},
\label{eq:GWs_spectrum_2a}
\end{equation}
%
where $\Omega_{\rm r_0}=2.47\times10^{-5}h^{-2}$ represents the current energy density of radiation.
In addition, $f_e=4.3\times 10^{8}~({\rm Hz})$, $f_{\rm re}$, $f_{\rm eq}=1.7\times 10^{-17}~({\rm Hz})$ and $f_*=2.4 \times 10^{-16}~({\rm Hz})$ denote the frequencies of GWs at the end of inflation, end of reheating, matter-radiation equality, and the CMB pivot scale $k_\ast=0.05~{\rm Mpc^{-1}}$, respectively. Note that in Eqs. (\ref{eq:GW_spectrum_2b}) and (\ref{eq:GWs_spectrum_2a}), the quantities $\omega_{\rm re}$, $r$ and $n_{\rm t}=-r/8$ as shown in Figs. \ref{fig:r-ns} and \ref{fig:wre} can be expressed in terms of the $e$-fold number $N$ and the model parameter $\alpha$.

The frequency of GWs is related to the comoving wavenumber $k$ as
\begin{equation}\label{fk}
 f=\frac{1}{2 \pi} \left( \frac{k}{a_0} \right).
\end{equation}
Subsequently, it can be expressed in terms of temperature as follows \cite{Mishra:2021}
\begin{equation}
f(T)=7.36\times 10^{-8} {~\rm Hz} \left ( \frac{g_{0}^{\rm s}}{g_{\rm T}^{\rm s}} \right )^{\frac{1}{3}}\left ( \frac{g_{\rm T}}{90} \right )^{\frac{1}{2}}\left ( \frac{T}{\rm GeV} \right ) ,
\label{eq:GW_f_master}
\end{equation}
where $g_0^{\rm s}=3.94 $ and $g_{\rm T}^{\rm s}=106.75 $ represent the effective number of relativistic degrees of freedom in entropy at the present time and at a temperature $T$, respectively. Additionally,  $g_{\rm T}=106.75$  represents the effective number of relativistic degrees of freedom in energy.

\begin{figure}[H]
\begin{minipage}[b]{1\textwidth}
\vspace{-.1cm}
\centering
\subfigure[\label{Gw45}]{\includegraphics[width=.48\textwidth]%
{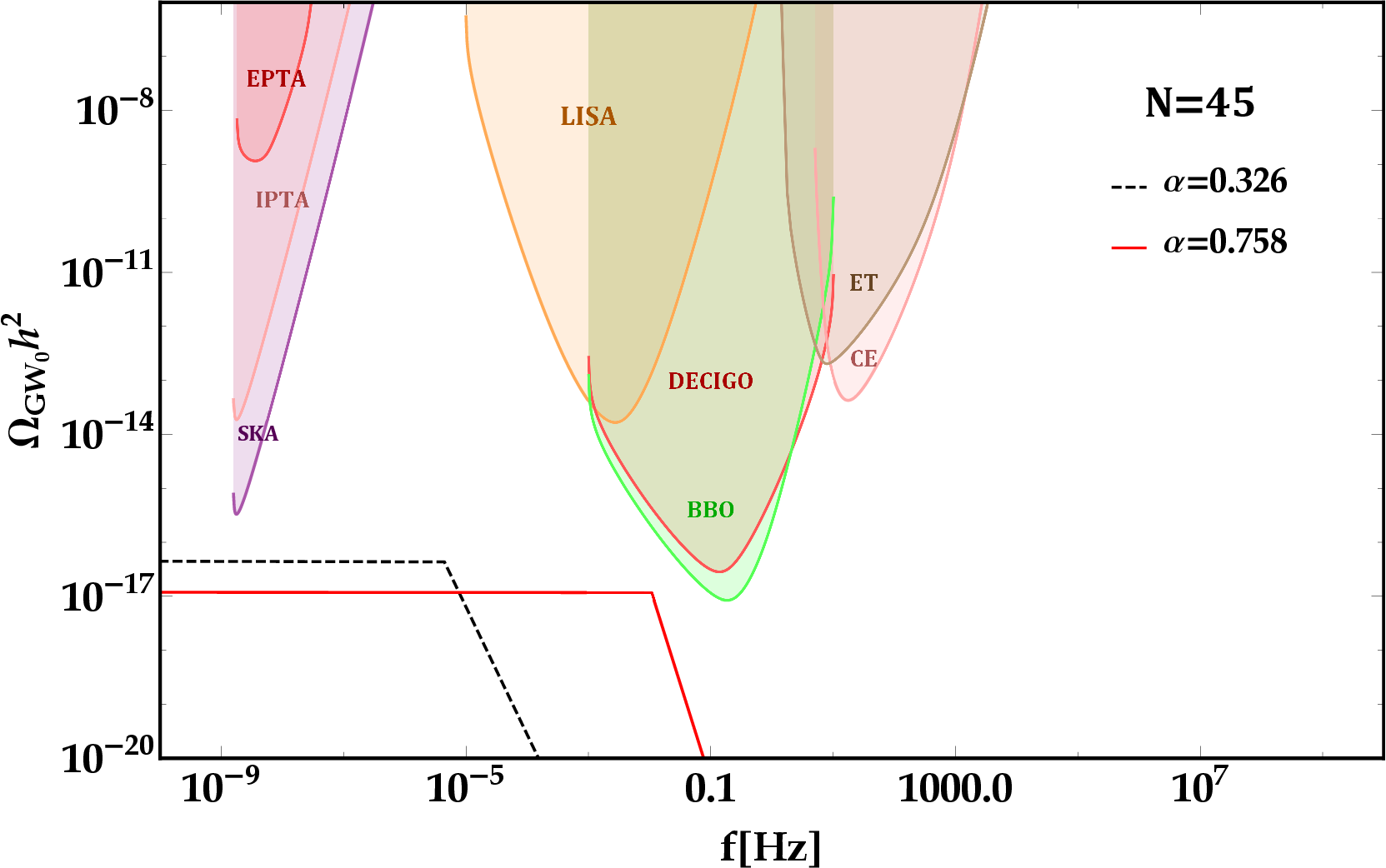}} \hspace{.1cm}
\centering
\subfigure[\label{GW46}]{\includegraphics[width=.48\textwidth]%
{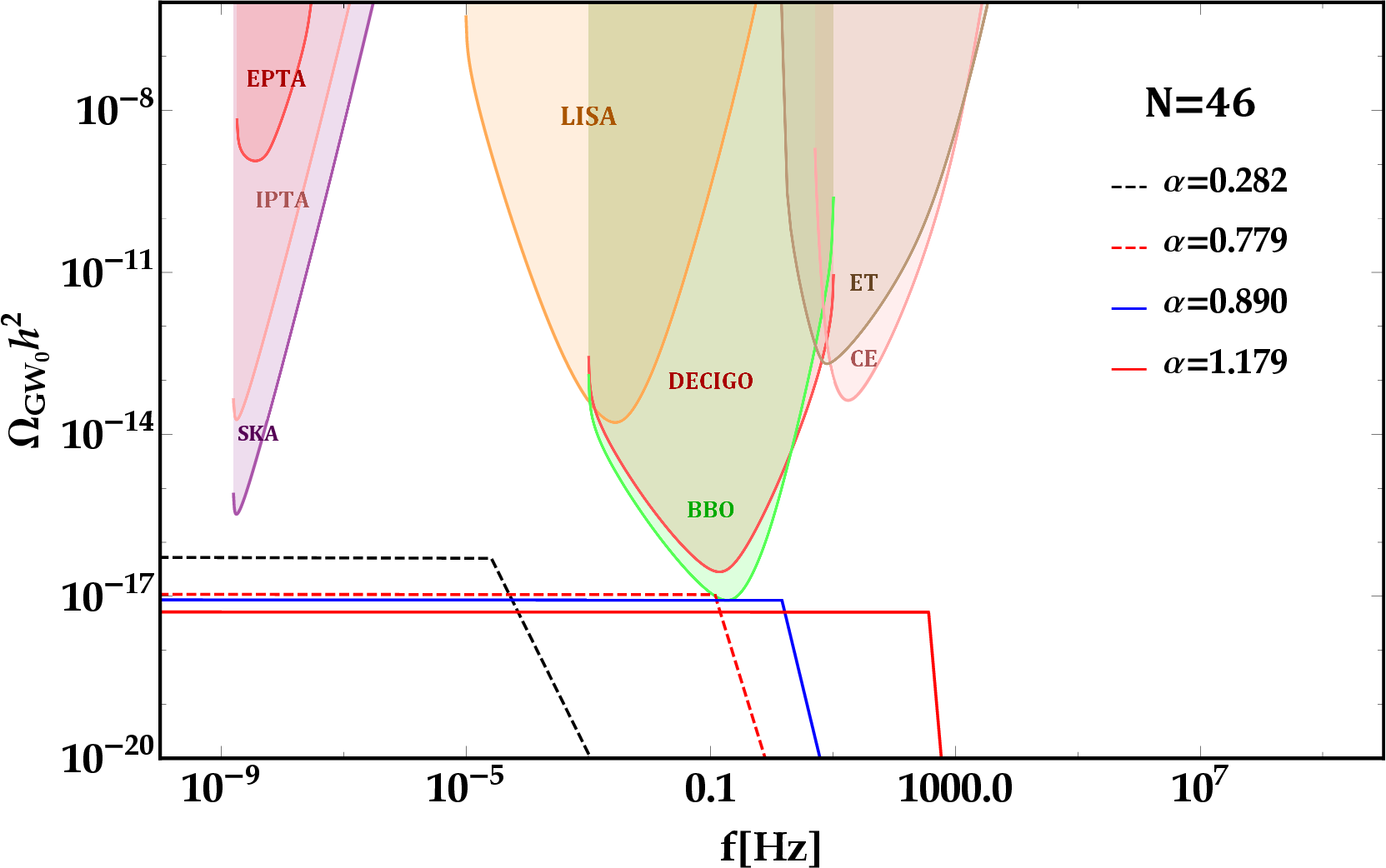}} \hspace{.1cm}
\centering
\subfigure[\label{GW48}]{\includegraphics[width=.48\textwidth]%
{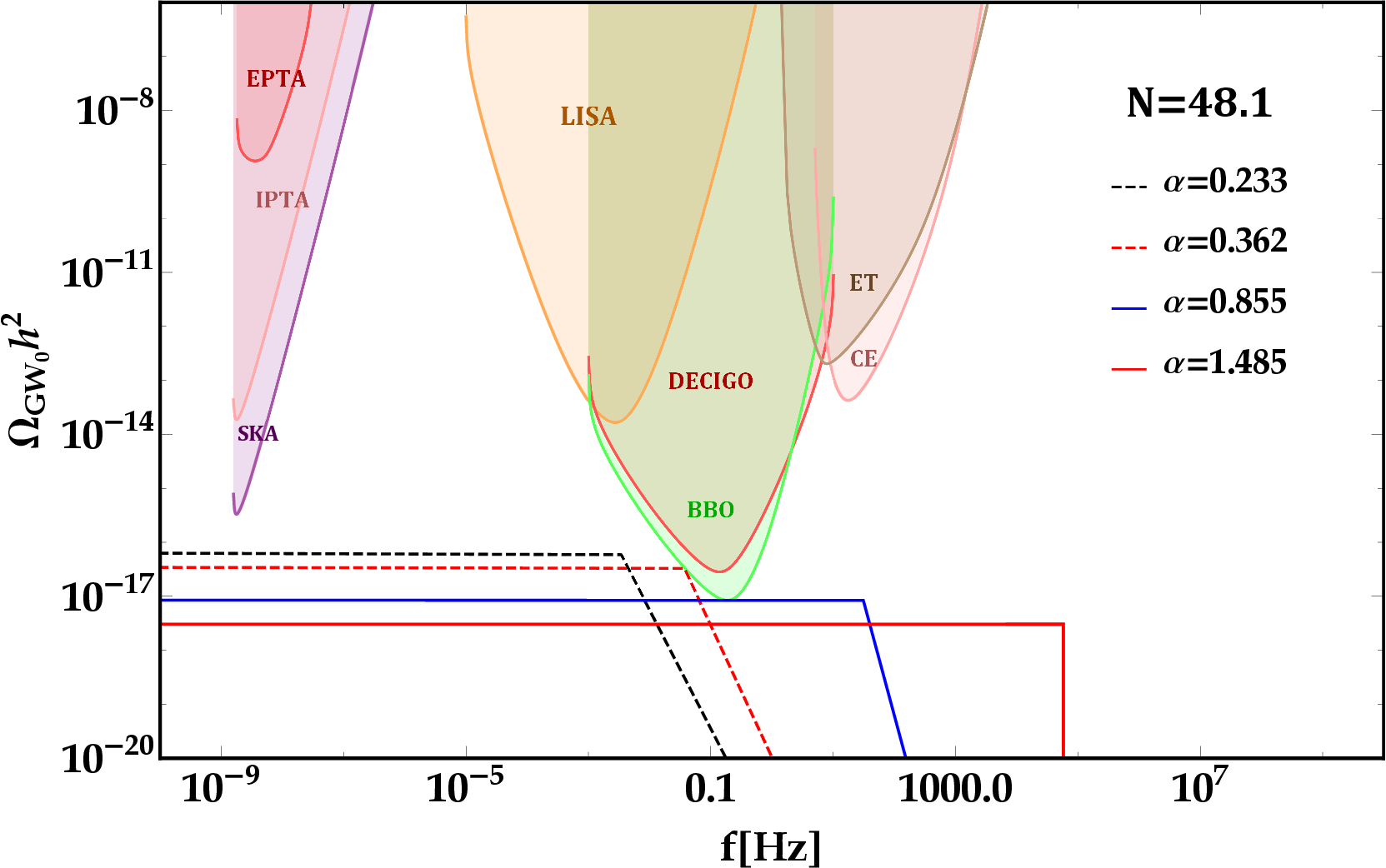}} \hspace{.1cm}
\centering
\subfigure[\label{GW50}]{\includegraphics[width=.48\textwidth]%
{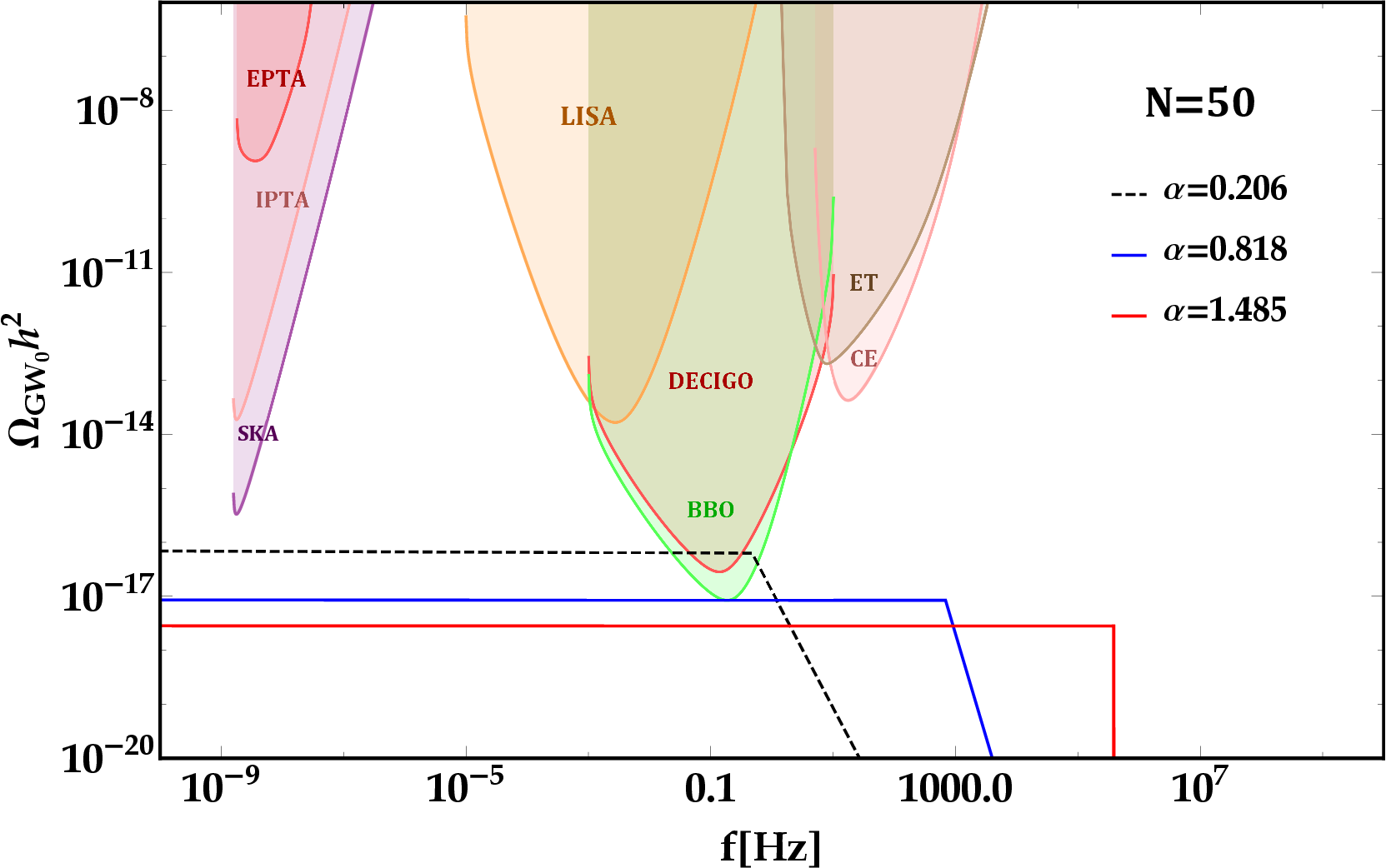}} \hspace{.1cm}
\centering
\subfigure[\label{GW55}]{\includegraphics[width=.48\textwidth]%
{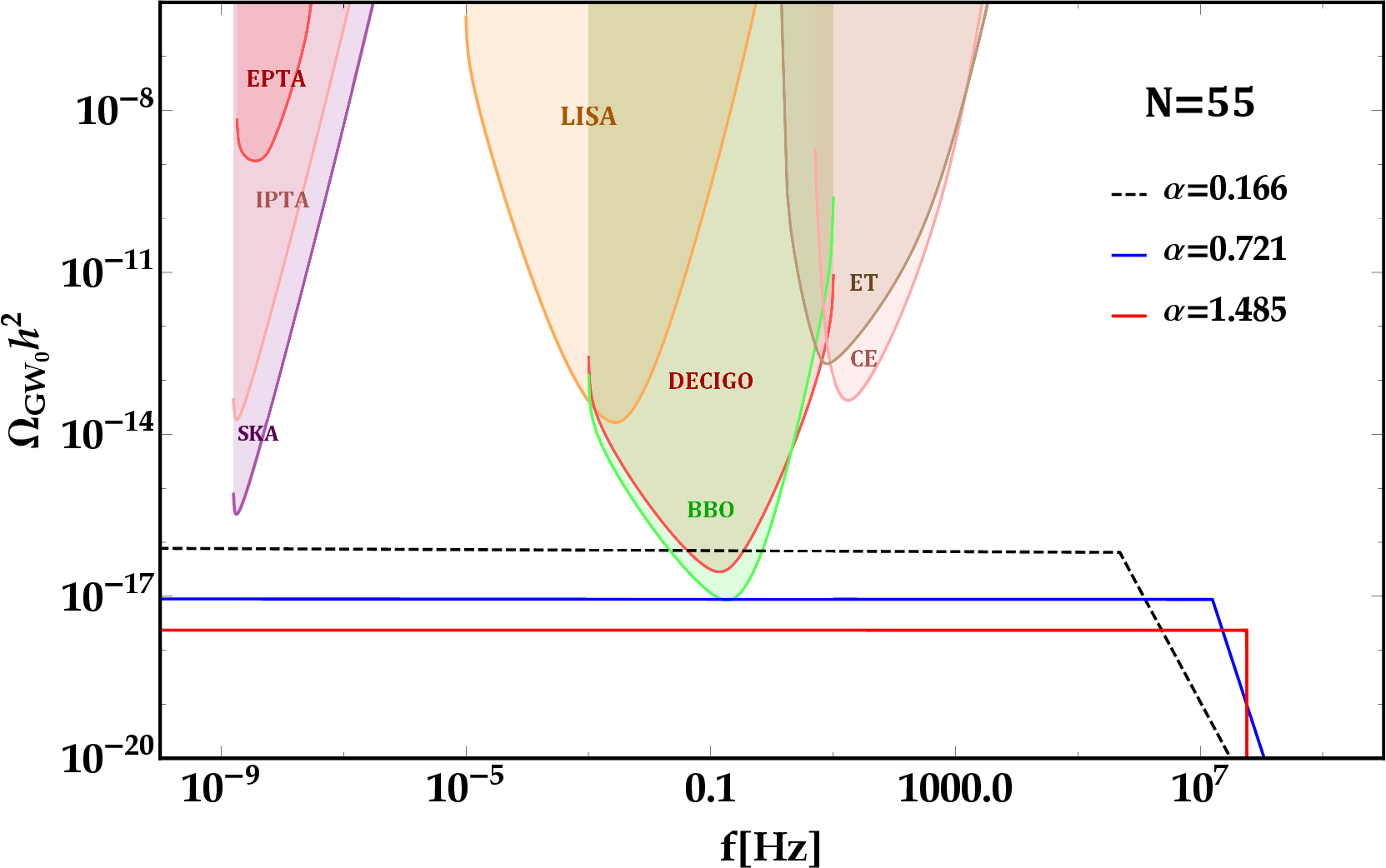}} \hspace{.1cm}
\centering
\subfigure[\label{GW56}]{\includegraphics[width=.48\textwidth]%
{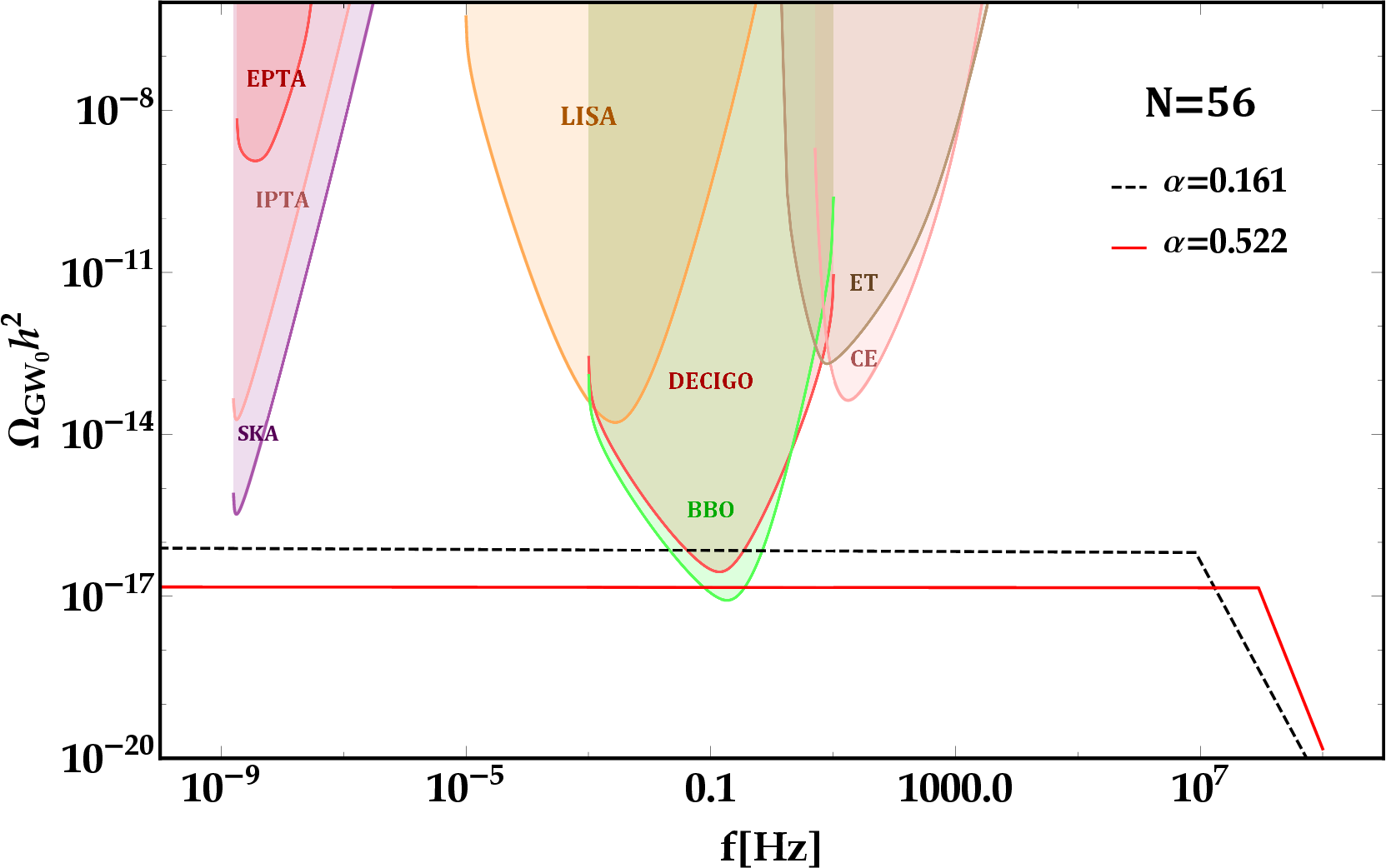}} \hspace{.1cm}
\end{minipage}
\caption{Energy density spectrum of relic GWs as a function of frequency for  $e$-fold numbers $N=(45,46,48.1,50,55,56)$ with different values of $\alpha$. The colored regions represent the sensitivity ranges of  GW detectors, including
BBO \cite{Yagi:2011BBODECIGO,Yagi:2017BBODECIGO,Harry:2006BBO,Crowder:2005BBO,Corbin:2006BBO},
DECIGO \cite{Yagi:2011BBODECIGO,Yagi:2017BBODECIGO,Seto:2001DECIGO,Kawamura:2006DECIGO,Kawamura:2011DECIGO},
LISA \cite{lisa,lisa-a},
SKA \cite{ska,skaCarilli:2004,skaWeltman:2020},
PTA \cite{epta1:add,epta2:add,epta3:add,epta4:add,epta5:add},
CE \cite{CE1,CE2}, and
ET \cite{ET1,ET2,ET3}.
The black (red) line in each graph corresponds to spectrum for the minimum (maximum) allowed values of $\alpha$, as listed in the third column in Table \ref{tabTre} (Planck 2018 + BK18 data).
The breaking point in the spectrum corresponds to the reheating frequency $f_{\rm re}$, which is linked to the reheating temperature $T_{\rm re}$ listed in Table \ref{tab GW1}.
}
\label{fig:GW}
\end{figure}

\begin{figure}[H]
\begin{minipage}[b]{1\textwidth}
\vspace{-.1cm}
\centering
\subfigure[\label{GW54ACT}]{\includegraphics[width=.48\textwidth]%
{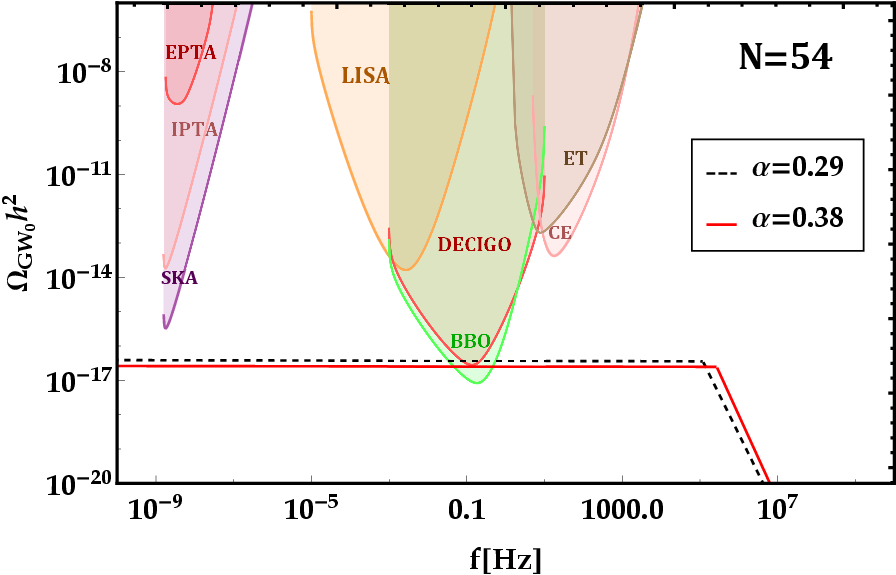}} \hspace{.1cm}
\centering
\subfigure[\label{GW55ACT}]{\includegraphics[width=.48\textwidth]%
{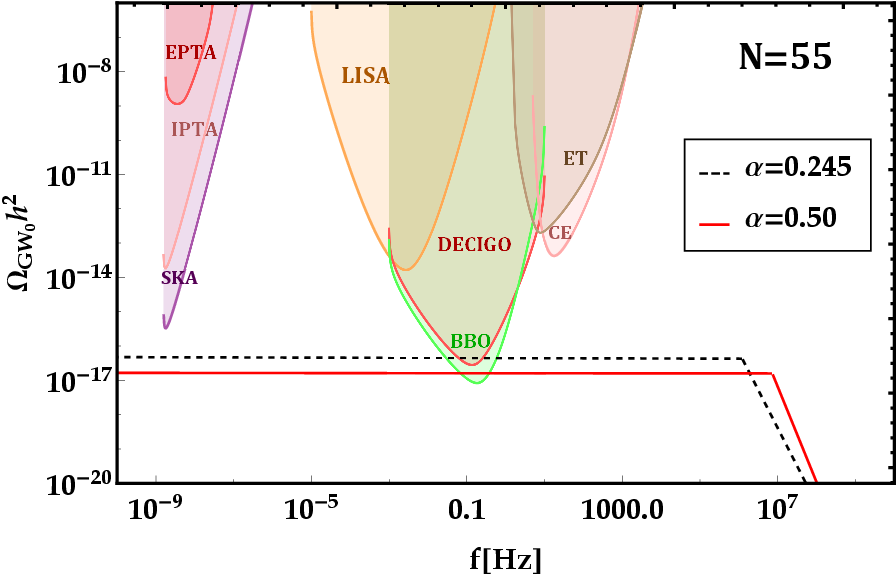}} \hspace{.1cm}
\centering
\subfigure[\label{GW56ACT}]{\includegraphics[width=.48\textwidth]%
{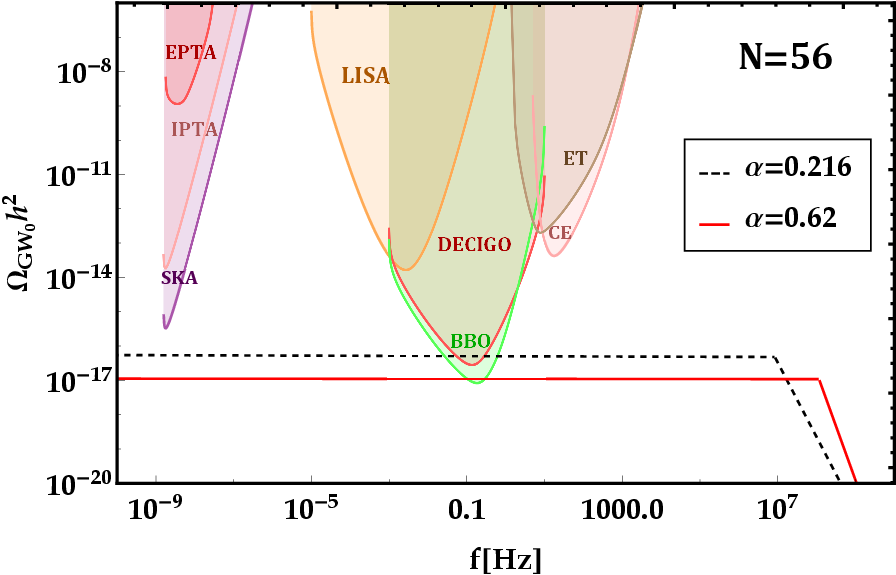}} \hspace{.1cm}
\end{minipage}
\caption{Energy density spectrum of relic GWs as a function of frequency for $e$-fold numbers $N=(54,55,56)$ with different values of $\alpha$. The colored regions correspond to the sensitivity domains of GW detectors, as in Fig. \ref{fig:GW}.
The black (red) lines represent the spectra for the minimum (maximum) allowed values of $\alpha$, according to the third column of Table \ref{tabTreACT} (Planck 2018 + BK18 + ACT data).
The break in each spectrum corresponds to the reheating frequency $f_{\rm re}$, which is linked to the reheating temperature $T_{\rm re}$ listed in Table \ref{tab GW1ACT}.
}
\label{fig:GWACT}
\end{figure}

To evaluate the detectability  of the relic GWs predicted by the present model, we compute their current density spectra for the allowed  ranges of $\alpha$ at specific numbers of $e$-fold, $N=(45,46,48.1,50,55,56)$ according to Table \ref{tabTre}, and $N=(54,55,56)$ according to Table \ref{tabTreACT}.
The resulting spectra are compared with the sensitivity ranges of GW detectors, as shown in Figs. \ref{fig:GW} and \ref{fig:GWACT}. In these figures: (i) the colored zones illustrate the sensitivity domains of various GW detectors,
BBO \cite{Yagi:2011BBODECIGO,Yagi:2017BBODECIGO,Harry:2006BBO,Crowder:2005BBO,Corbin:2006BBO},
DECIGO \cite{Yagi:2011BBODECIGO,Yagi:2017BBODECIGO,Seto:2001DECIGO,Kawamura:2006DECIGO,Kawamura:2011DECIGO},
LISA \cite{lisa,lisa-a},
SKA \cite{ska,skaCarilli:2004,skaWeltman:2020},
PTA \cite{epta1:add,epta2:add,epta3:add,epta4:add,epta5:add}, CE \cite{CE1,CE2}, and
ET \cite{ET1,ET2,ET3}; (ii)
the black (red) curves correspond to the minimum (maximum) allowed values of $\alpha$ for a given inflationary duration $N$ as listed in Tables \ref{tabTre} and \ref{tabTreACT}; (iii) the break in each spectrum corresponds to the frequency at the end of the reheating, $f_{\rm re}=f(T=T_{\rm re})$, which from Eq. (\ref{eq:GW_f_master}) is directly related to the reheating temperature $T_{\rm re}$ in Eq. (\ref{eq:Tre}) (see Tables \ref{tab GW1} and \ref{tab GW1ACT}).

In Fig. \ref{fig:GW} (Planck 2018 + BK18 data): (i) for $N=45$, the generated GWs lie below the sensitivity of GW observatories (Fig. \ref{Gw45}); (ii) for $N=(46,48.1,50,55$),
the ranges of $\alpha$ in Table \ref{tabTre} must be slightly adjusted to produce detectable signals (Figs. \ref{GW46}-\ref{GW55}, last column of Table \ref{tabTre}); (iii) for $N=46$, detectable GWs are produced for  $ 0.779\leq\alpha\leq 0.890$ (95\% CL)  (Fig. \ref{GW46}); (iv) for $N=48.1$,  the 95\% CL range is modified to $ 0.362\leq\alpha\leq 0.855$ to produce detectable GWs (Fig. \ref{GW48}, last column of Table \ref{tabTre});
(v) for $N=50$, GWs corresponding to the lower bound of $\alpha$ are detectable by
BBO \cite{Yagi:2011BBODECIGO,Yagi:2017BBODECIGO,Harry:2006BBO,Crowder:2005BBO,Corbin:2006BBO} and DECIGO \cite{Yagi:2011BBODECIGO,Yagi:2017BBODECIGO,Seto:2001DECIGO,Kawamura:2006DECIGO,
Kawamura:2011DECIGO} (black curve in Fig. \ref{GW50}), while the upper limit must be modified to $0.818$ (95\% CL) for detectablility  (blue curve in Fig. \ref{GW50});
(vi) for $N=55$, the upper bound changes from  $\alpha=1.485$ to $0.721$ (95\% and 68\% CL) (blue curve in Fig. \ref{GW55}); and (vii) for
 $N=56$, the previously obtained range of $\alpha$ in the third column of Table \ref{tabTre} already yields a detectable GW spectrum (Fig. \ref{GW56}).

In Fig. \ref{fig:GWACT} (Planck 2018 + BK18 + ACT data): (i) for $N=54$, GWs associated with both the lower and upper bounds of $\alpha$ are detectable by BBO \cite{Yagi:2011BBODECIGO,Yagi:2017BBODECIGO,Harry:2006BBO,Crowder:2005BBO,Corbin:2006BBO} and DECIGO \cite{Yagi:2011BBODECIGO,Yagi:2017BBODECIGO,Seto:2001DECIGO,Kawamura:2006DECIGO,
Kawamura:2011DECIGO} (Fig. \ref{GW54ACT}); (ii) for $N=55$ (N=56), the spectra corresponding to the upper bounds $\alpha=0.50$ ($\alpha=0.62$) are detectable by BBO, while those corresponding to the lower bounds $\alpha=0.245$ ($\alpha=0.216$) are detectable by DECIGO (Fig \ref{GW55ACT}).

Overall, the relic GW constraints indicate that the minimum permissible inflationary duration at the 95\% CL is  $N=46$ with $0.779\leq\alpha\leq0.890$ for the  Planck 2018 + BK18 dataset. The permitted ranges of the inflationary duration are  $46\leq N\leq 56$ (95\% CL) and $48.1\leq N\leq 56$ (68\% CL) for Planck + BK18, while inclusion of the  ACT data tighten these to $54\leq N\leq 56$ (95\% CL) (see the last columns of Tables \ref{tabTre} and \ref{tabTreACT}).



\begin{table}[H]
  \centering
  \caption{Reheating temperature $T_{\rm re}$, Eq. (\ref{eq:Tre}), and frequency $f_{\rm re}=f(T=T_{\rm re})$, Eq. (\ref{eq:GW_f_master}) for various numbers of $e$-fold  $N$ and values of $\alpha$ according to Table \ref{tabTre} (Planck 2018 + BK18 data). See also  Fig. \ref{Tre_a}.}
\resizebox{.6\textwidth}{!}{\scriptsize
\begin{tabular}{cccc}
\thickhline
$N$\qquad  &
\qquad$\alpha$\qquad &
\qquad $T_{\rm re}/{\rm GeV} $\qquad &
\qquad $f_{\rm re}/{\rm Hz} $\qquad  \\
\thickhline
\quad  \multirow{2}{*}{45} \quad\qquad & 
\qquad$ 0.326 $ \qquad  &
\qquad $1.67\times 10^2 $\qquad &
\qquad $ 4.43\times10^{-6}$  \\
\quad &
\qquad$ 0.758 $ \qquad  &
\qquad $ 4.04\times10^{5}  $\qquad &
\qquad $ 1.08\times10^{-2} $  \\
\hline
\quad $\multirow{4}{*}{46} $\quad\qquad  & 
\qquad$ 0.282 $ \qquad &
\qquad $ 9.73\times10^{2} $ &
\qquad $ 2.59\times10^{-5} $ \\
\quad &
\qquad$ 0.779 $ \qquad &
\qquad $ 4.36\times10^{6} $ &
\qquad $ 1.16\times10^{-1} $ \\
\quad &
\qquad$ 0.890 $ \qquad &
\qquad $ 5.45 \times10^{7} $ &
\qquad $ 1.45 $  \\
\quad &
\qquad$ 1.179 $ \qquad &
\qquad $ 1.36 \times10^{10} $ &
\qquad $ 3.63\times10^{2} $  \\
\hline
\quad $\multirow{4}{*}{48.1} $\quad\qquad  & 
\qquad$ 0.233 $ \qquad &
\qquad $ 1.28\times10^{5} $ &
\qquad $ 3.42\times10^{-3} $ \\
\quad &
\qquad$ 0.362 $ \qquad &
\qquad $ 1.38\times10^{6} $ &
\qquad $ 3.68\times10^{-2} $ \\
\quad &
\qquad$ 0.855 $ \qquad &
\qquad $ 1.17 \times10^{9} $ &
\qquad $ 3.12 \times10$  \\
\quad &
\qquad$ 1.485 $ \qquad &
\qquad $ 2.17 \times10^{12} $ &
\qquad $ 5.79\times10^{4} $  \\
\hline
\quad $\multirow{3}{*}{50} $\quad\qquad  & 
\qquad$ 0.206 $ \qquad &
\qquad $ 1.75\times10^{7} $ &
\qquad $ 4.68\times10^{-1} $ \\
\quad &
\qquad$ 0.818 $ \qquad &
\qquad $ 2.57 \times10^{10} $ &
\qquad $ 6.87\times10^{2} $  \\
\quad &
\qquad$ 1.485 $ \qquad &
\qquad $ 1.45 \times10^{13} $ &
\qquad $ 3.87\times10^{5} $  \\
\hline
 \quad $\multirow{3}{*}{55} $\quad\qquad   & 
\qquad$ 0.166 $ \qquad &
\qquad $ 1.81\times10^{13} $ &
\qquad $ 4.84\times10^{5} $ \\
\quad  &
\qquad$ 0.721 $ \qquad &
\qquad $ 5.92 \times10^{14} $ &
\qquad $ 1.58\times10^{7} $  \\
\quad &
\qquad$ 1.485 $ \qquad &
\qquad $ 2.16 \times10^{15} $ &
\qquad $ 5.77\times10^{7} $  \\
\hline
\quad $\multirow{3}{*}{56} $\quad\qquad   & 
\qquad$ 0.161 $ \qquad &
\qquad $ 3.14\times10^{14} $ &
\qquad $ 8.38\times10^{6} $ \\
\quad  &
\qquad$ 0.217 $ \qquad &
\qquad $ 5.87 \times10^{14} $ &
\qquad $ 1.57\times10^{7} $  \\
\quad &
\qquad$ 0.522 $ \qquad &
\qquad $ 3.35 \times10^{15} $ &
\qquad $ 8.94 \times10^{7} $  \\
\thickhline
\end{tabular}}
 \label{tab GW1}
\end{table}

\begin{table}[H]
  \centering
  \caption{Reheating temperature $T_{\rm re}$, Eq. (\ref{eq:Tre}), and frequency $f_{\rm re}=f(T=T_{\rm re})$, Eq. (\ref{eq:GW_f_master}) for different numbers of $e$-fold $N$  and values of $\alpha$, obtained by incorporating the ACT data (Planck 2018 + BK18 + ACT). See also  Fig. \ref{Tre_aACT}.}
\resizebox{.6\textwidth}{!}{\scriptsize
\begin{tabular}{cccc}
\thickhline
$N$\qquad  &
\qquad$\alpha$\qquad &
\qquad $T_{\rm re}/{\rm GeV} $\qquad &
\qquad $f_{\rm re}/{\rm Hz} $\qquad  \\
\thickhline
\quad $\multirow{2}{*}{54} $\quad\qquad  & 
\qquad$ 0.29 $ \qquad &
\qquad $ 4.44\times10^{12} $ &
\qquad $ 1.18\times10^{5} $ \\
\quad &
\qquad$ 0.38 $ \qquad &
\qquad $ 1.01 \times10^{13} $ &
\qquad $ 2.69\times10^{5} $  \\
\hline
 \quad $\multirow{2}{*}{55} $\quad\qquad   & 
\qquad$ 0.245 $ \qquad &
\qquad $ 4.57\times10^{13} $ &
\qquad $ 1.22\times10^{6} $ \\
\quad  &
\qquad$ 0.50 $ \qquad &
\qquad $ 2.72 \times10^{14} $ &
\qquad $ 7.25\times10^{7} $  \\
\hline
\quad $\multirow{2}{*}{56} $\quad\qquad   & 
\qquad$ 0.216 $ \qquad &
\qquad $ 3.15\times10^{14} $ &
\qquad $ 8.42\times10^{6} $ \\
\quad  &
\qquad$ 0.62 $ \qquad &
\qquad $ 4.32 \times10^{15} $ &
\qquad $ 1.15\times10^{8} $  \\
\thickhline
\end{tabular}}
 \label{tab GW1ACT}
\end{table}

\section{Conclusions}\label{sec5}
Within the framework of Einstein gravity, we studied an inflationary model driven by the mutated hilltop potential, a flexible extension of the standard hilltop model. Our analysis demonstrates that the scalar spectral index $n_{\rm s}$ and tensor-to-scalar ratio $r$ predicted by this model remain compatible with the combined constraints from Planck 2018 and BICEP/Keck 2018. The inclusion of the ACT data further refines the allowed parameter space which yields tighter bounds on $N$ and $\alpha$. We further constrained the parameter space by incorporating the effects of reheating and RD era. Specifically, we analyzed the dependence of the reheating equation of state parameter $\omega_{\rm re}$, reheating duration $N_{\rm re}$, reheating temperature $T_{\rm re}$, and  RD duration $N_{\rm rd}$ on the parameter $\alpha$. By combining these constraints with CMB and relic GWs, we obtained the following key results:
\begin{itemize}
  \item The mutated hilltop inflationary model is in good agreement with the latest observational  $r-n_{\rm s}$ data from Planck and BICEP/Keck 2018, and remains consistent when ACT data are incorporated, for specific ranges of $N$ and $\alpha$ (Fig. \ref{fig:r-ns} and Tables \ref{tabTre}-\ref{tabTreACT}).
  \item Planck 2018 + BK18 constraints impose a minimum inflationary duration of $N=44.4$ (95\%  CL) with $0.372\leq \alpha\leq 0.642$ and $N=48.1$ (68\% CL) with $0.410\leq \alpha\leq 0.490$. Incorporating the ACT data shifts these bounds to $N=54$ (95\%  CL) with $0.29\leq \alpha\leq 0.38$ (See Tables \ref{tabTre}-\ref{tabTreACT}).
  \item The reheating equation of state parameter $\omega_{\rm reh}$ varies with $\alpha$  and is constrained to the range $-1/3\leq\omega_{\rm reh}<0$. This imposes an upper bound $\alpha\leq 1.485$ (Fig. \ref{fig:wre}).
  \item By combining CMB and reheating constraints, the observable inflationary duration is bounded to $44.4\leq N\leq 56$ for $0.161\leq\alpha \leq 1.485$ (95\% CL)  and $48.1\leq N\leq 56$ for $0.217\leq\alpha \leq 1.485$ (68\% CL), based on Planck 2018 + BK18 data.  When ACT data are included, the bounds tighten to $54\leq N\leq 56$ for $0.29\leq\alpha \leq 0.62$ (95\% CL). The upper bound $N=56$ arises from the model independent reheating condition $N_{\rm re}\geq 0$ (Figs. \ref{Nre_a} and \ref{Nre_aACT}).
   \item The constraint the RD era duration $N_{\rm rd}$ does not significantly alter the bounds on  $N$  and $\alpha$ deduced from the CMB and reheating analyses.
   \item When combining  CMB (Planck 2018 + BK18), reheating, RD era and relic GW constraints, the observable inflationary duration is further refined to $46\leq N\leq 56$ for $0.161\leq\alpha \leq 0.890$ (95\% CL) and $48.1\leq N\leq 56$ for $0.217\leq\alpha \leq 0.815$ (68\% CL) (see the last column of Table \ref{tabTre}). The lower bound $N=46$ and the upper bounds on $\alpha$ are determined by the detectability of the relic GWs spectra (Fig. \ref{fig:GW}).
   \item When ACT data are included (Planck 2018 + BK18 + ACT), the combination of CMB, reheating, and RD constraints with relic GWs does not further modify the allowed parameter space, which remains  $54\leq N\leq 56$ for $0.29\leq\alpha \leq 0.62$ (95\% CL) (Table \ref{tabTreACT}, Fig. \ref{fig:GWACT}).
  \item The predicted relic GW spectra corresponding to $46 \leq N \leq 56$ (Planck 2018 + BK18) and  $54 \leq N \leq 56$ (Planck 2018 + BK18 + ACT) fall within the sensitivity domains of various GW detectors and provide a promising observational test for this class of inflationary models.
\end{itemize}
In summary, the mutated hilltop inflationary model remains fully consistent with current observational data for special ranges of $N$ and $\alpha$. The inclusion of ACT data significantly narrows the viable parameter space, while relic GW considerations introduce an additional, independent layer of constraint. This study highlights the importance of combining multiple observational and theoretical bounds to robustly constrain inflationary models and to enhance their predictive power in the era of precision cosmology.



\end{document}